\documentclass[aps,prb,twocolumn,floatfix,superscriptaddress,longbibliography,nofootinbib,10pt]{revtex4-2}

\usepackage{graphicx, color, soul}
\usepackage{amsmath, amsfonts, amssymb, mathrsfs, dsfont, mathtools}

\usepackage[dvipsnames, pdftex, table]{xcolor}
\usepackage[normalem]{ulem} % needed for \scrap-command
\usepackage[mathscr]{eucal}
\usepackage{bm}
\usepackage{tabularx, booktabs}
\usepackage{multirow}
\usepackage{physics}
\usepackage{bbold}
\usepackage{comment}
\usepackage{wasysym}
\usepackage{tikz}    % For drawing shapes
\usepackage{stackengine}% for stacking symbols
\usepackage{csquotes}
\usepackage[colorlinks, breaklinks, 
            linkcolor=Blue,
            citecolor=Blue,
            urlcolor=Blue]{hyperref}
            
\usepackage[all]{hypcap} 
\usepackage{enumitem}
\usepackage{placeins}
\DeclareMathOperator{\sgn}{sgn}

\begin{document}
	
\title{Spin Chern phases and persistent spin texture in a quasi-2D SSH model}
	
\author{Hemant K Sharma}
\affiliation{Department of Physics, National Institute of Science Education and Research, Jatni, 752050, India}
\affiliation{Homi Bhabha National Institute, Training School Complex, Anushakti Nagar, Mumbai 400094, India}
\author{Saptarshi Mandal}
\affiliation{Institute of Physics, Sachivalaya Marg, Bhubaneswar-751005, India}
\affiliation{Homi Bhabha National Institute, Training School Complex, Anushakti Nagar, Mumbai 400094, India}
\author{Kush Saha}
\affiliation{Max-Planck Institute for the Physics of Complex Systems, Noethnitzer Str. 38, 01187, Dresden, Germany}
\affiliation{Department of Physics, National Institute of Science Education and Research, Jatni, 752050, India}
\affiliation{Homi Bhabha National Institute, Training School Complex, Anushakti Nagar, Mumbai 400094, India}
	
\begin{abstract}
%We construct a quasi-two-dimensional Su-Schrieffer-Heeger model (SSH)-like model and uncover a rich set of topological phases with nontrivial spin textures in the presence of complex hopping and spin–orbit coupling. Despite its simple structure, the combined effect of complex hopping and spin–orbit interaction  gives rise not only to the conventional quantum anomalous Hall insulator (QAHI) phase, but also to  distinct combinations of quantum anomalous spin Hall insulator (QASHI) phase.  We further show that the bulk bands of this model can host persistent spin textures, whose emergence is controlled by the relative strength of the nearest- and next-nearest-neighbor complex hopping. We develop a low-energy continuum description that captures the emergence of these phases and clarifies the origin of the persistent spin structures. Interestingly, the resulting spin textures closely resemble those commonly found in conventional semiconductor systems with trivial band topology, yet they arise in a nontrivial topological setting, enabled by carefully engineered hopping patterns.
We construct a quasi-two-dimensional Su-Schrieffer-Heeger model (SSH)-like model and uncover a rich set of topological phases with nontrivial spin textures in the presence of complex hopping and spin–orbit coupling. Despite its simple structure, the combined effect of complex hopping and spin–orbit interaction  gives rise not only to the conventional quantum anomalous Hall insulating (QAHI) phase, but also to  distinct combinations of spin Chern phases, namely quantum anomalous spin Hall insulating (QASHI) phase.  Furthermore, we demonstrate that the bulk bands of this model can host persistent spin textures, whose formation and stability are governed by the relative strengths of nearest- and next-nearest-neighbor complex hopping. To elucidate the underlying mechanisms, we develop a low-energy continuum theory that captures the emergence of these topological phases and clarifies the origin of the persistent spin textures. Interestingly, the resulting spin textures closely resemble those typically observed in conventional semiconductor systems with topologically trivial band structures. However, in our case, they emerge within a nontrivial topological framework, enabled by carefully engineered hopping patterns that intertwine lattice geometry, complex hopping, and spin–orbit coupling

	\end{abstract}
	
	\maketitle
	
\section{Introduction}
\begin{figure}
\includegraphics[width=1.0\linewidth]{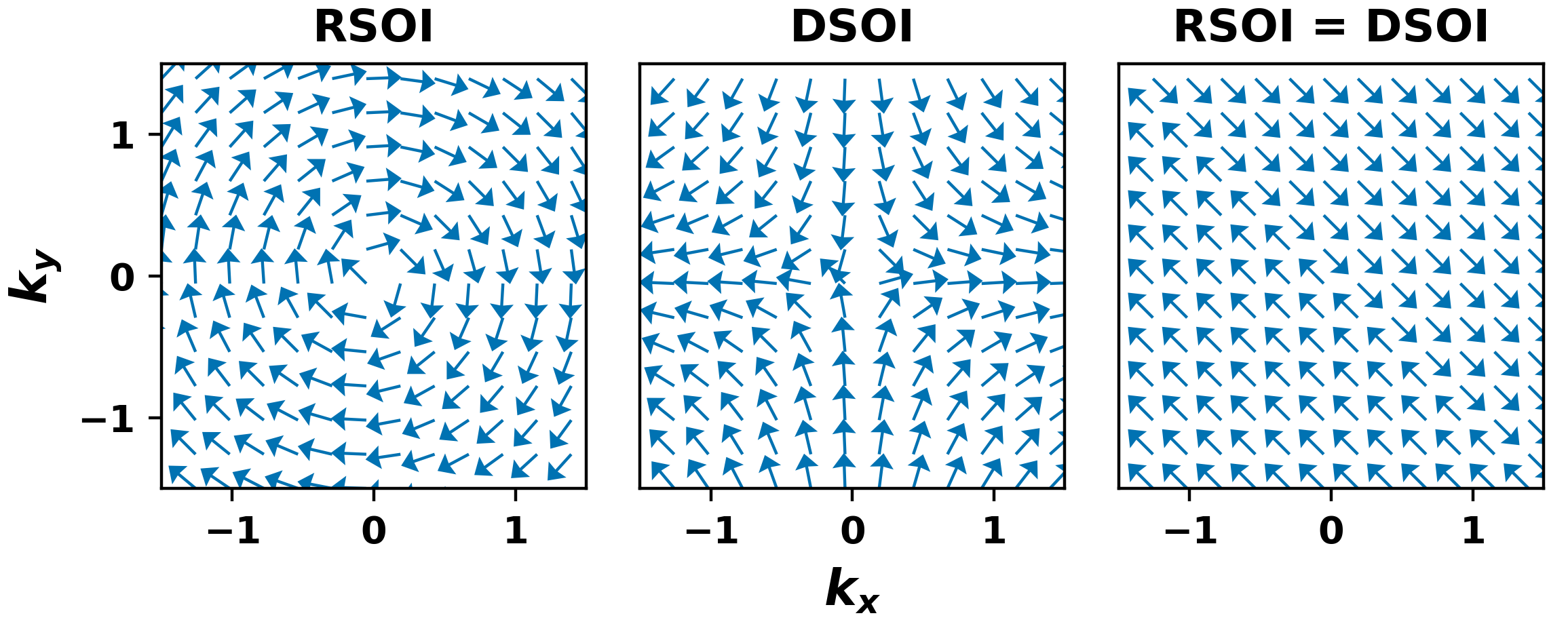}
\caption{Spin texture of Rashba SOC with effective momentum-dependent vector field, ${\bf h}_{\rm RSOC}=(k_y,-k_x)$ (a), Dresselhauss SOC with effective vector field, ${\bf h}_{\rm DSOC}=(k_x, -k_y)$, (b) and both the SOC with equal strength leading to an effective field, ${\bf h}_{\rm DSOC+\rm RSOC}=(k_y+kx, -k_x-k_y)$ (c). Evidently, we see unidirectional spin texture when both SOCs are of equal strength.}
\label{fig1:SSOCDRSOC}
\end{figure}

 The coupling between spin and orbital degrees of freedom, known as spin–orbit coupling (SOC), plays a central role in determining novel physical properties of materials, ranging from conventional semiconductors to topological quantum materials \cite{Hasan2010,QI2011,Tokura2019}. Owing to spin–momentum locking, SOC can generate a variety of momentum-dependent spin textures such as helical, vortex-like, and skyrmionic patterns \cite{Hsieh2009,Murakami2004}. These spin textures reflect the underlying symmetries of the Hamiltonian and strongly influence spin transport, magnetoelectric responses, and the topological characteristics of a material\cite{PhysRevB.109.085142, Maass2016, PhysRevB.111.115158, Yaji2017SpinDependentQuantum, 
Xie2014OrbitalSelective, xt23-9pnv}.
In addition to these conventional textures, a momentum-independent spin configuration, known as a persistent spin texture (PST), can emerge in systems with SOC when bulk and interfacial inversion symmetries are broken together, or due to specific nonsymmorphic space-group symmetries \cite{tao_etal}. Among the various types of spin textures, PSTs have attracted particular attention because of their potential applications in spintronic devices \cite{Schliemann2017,Koralek2009}. Unlike ordinary spin textures, which decay due to spin relaxation mechanisms, a PST exhibits a unidirectional spin alignment that is preserved across the entire Brillouin zone.  In certain two-dimensional electron gases (2DEGs) \cite{Bernevig2006,Cartoixa2005} this robustness originates from an emergent SU(2) spin-rotation symmetry, which appears when Rashba and Dresselhaus spin–orbit couplings are exactly balanced (see Fig.~\ref{fig1:SSOCDRSOC}). As a result, the spin configuration is protected against D’yakonov–Perel’ spin relaxation \cite{Dyakonov1972}, leading to exceptionally long spin lifetimes which is an essential feature for spin-based information processing \cite{Dettwiler2017,Kunihashi2016}. Beyond their potential technological applications, PSTs provide a clear manifestation of the interplay between crystalline symmetries and SOC in generating robust, momentum-space spin coherence.

While semiconductors are well-known candidates for hosting persistent spin textures (PST), primarily due to Rashba and Dresselhaus spin–orbit coupling arising from surface or interface asymmetry under various point groups or due to certain nonsymmorphic symmetries\cite{Schliemann2017,Koralek2009,tao_etal,berkay,junyi}, topological materials have been mostly overlooked as potential PST platforms\cite{PhysRevB.110.195142}. This is, in part, because topological systems typically exhibit strong spin–orbit coupling, which in turn generate strongly momentum-dependent spin textures. On the other hand, only a limited number of them possess the specific nonsymmorphic crystalline symmetries required to protect PST\cite{PhysRevB.90.085304}. Recent  theoretical reports on unconventional quantum Hall phase featuring canted persistent spin structure in WTe$_2$ and tungsten-based transition metal dichalcogenides (TMDCs)\cite{jose_prl2020,moh_PRB2025} highlight that robust, symmetry-protected spin coherence can indeed emerge in topological settings. These findings not only challenge the prevailing assumption that strong spin–orbit coupling precludes PST, but also open a new direction for topological materials as potential platforms for persistent spin textures.

In view of the above, we construct a spinful quasi-two dimensional SSH model and introduce spin-orbit coupling along with a few specific complex hopping. The inclusion of the spin-orbit coupling together with complex hopping drives the system into various spin-resolved topological phases. In addition to the standard \emph{quantum anomalous  Hall insulator} (QAHI) phase, which is characterized by $C_{\uparrow}=-1,C_{\downarrow}=-1$, we find \emph{quantum anomalous spin Hall insulator} (QASHI). The QASHI is characterised by  $C_{\uparrow}\ne C_{\downarrow}$. In particular, we find three distinct variant of QASH phases as ($C_{\uparrow}=-1, C_{\downarrow}=0$),   ($C_{\uparrow}=0, C_{\downarrow}=-1$)  and ($C_{\uparrow}=1, C_{\downarrow}=0$) depending on the values of complex hopping. These phases can dictate whether the edge transport is chiral, or spin-filtered. We then focus on the detailed structure of the spin texture in momentum space. Remarkably, we observe that in certain parameter regions, the spin orientations of the bulk bands become nearly uniform around some specific high symmetry $k-$points, forming PST of the bulk bands, as opposed to the standard topological materials. This PST arises not from a fine-tuned balance of SOC terms, as in conventional 2DEGs, or due to any space-group symmetries\cite{Dey2025CurrentInduced,rcg6-3gxl,PhysRevB.101.195418,PhysRevB.101.155410, PhysRevLett.125.216405, PhysRevMaterials.3.054407, tm58-lbdl, PhysRevB.110.235162, PhysRevB.99.125404, Tenzin2025PersistentSpin, PhysRevMaterials.6.094602,PhysRevB.104.115145}, rather due to presence of complex hopping together with the SOC terms. We provide a low-energy description of the distinct topological phases and elaborate on how the presence of complex hopping leads to the emergence of non-trivial spin textures. Our findings open up new pathways for designing low-dissipation spintronic devices based on lattice-engineered topological materials~\cite{Tokura2019,Ozawa2019,Mukherjee2020,Atala2013}.

\section{Model Construction}
\begin{figure}
\includegraphics[width=1.0\linewidth]{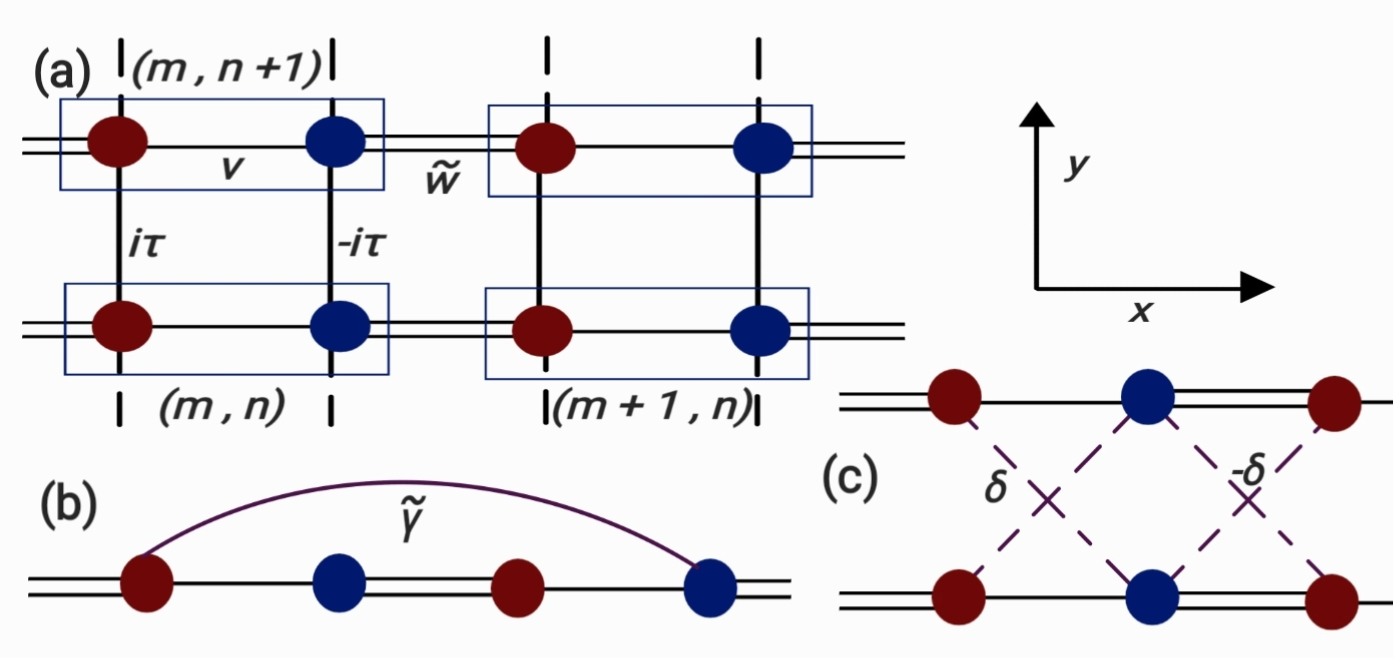}
\caption{Schematic illustration of the quasi two-dimensional spinful SSH model. One-dimensional SSH chains extend along the $x$ direction and are coupled along the $y$ direction to form a ladder-like 2D lattice geometry. The hopping $\gamma$ represents anisotropy between different sublattices. The interlayer hopping $\delta$ in (c) shows staggered pattern.}
\label{fig1:lattice}
\end{figure}

We begin with a spinful one-dimensional (1D) SSH chain oriented along $x$ direction. We then introduce next-to-next-nearest neighbor ($\gamma$) hopping between two-sublattices in the unit cells to obtain an extended 1D SSH model (see Fig.~\ref{fig1:lattice}b). To construct its two-dimensional counterpart, we stack these chains as ladders along the $y$ direction. We then incorporate equal but opposite in sign complex nearest-neighbor coupling ($i\tau$) between interlayer two equivalent sublattices. Additionally, we introduce staggered hopping ($\delta$) between identical sublattices as shown in  Fig.~\ref{fig1:lattice}c. The tight-binding Hamiltonian for this construction reads off {\textcolor{red}{,}}
\begin{eqnarray}
	\label{eqn:main_ham}
	H_0 = &&\sum_{m,n}\Big[
	{\psi}^\dagger_{m,n}h_0\bm{\psi}_{m,n}
	+ {\psi}^\dagger_{m,n}\,{\mathcal T_x} {\psi}_{m+1,n} \nonumber\\
	&&+ {\psi}^\dagger_{m,n}\,{\mathcal T_y} {\psi}_{m,n+1}
	+ {\psi}^\dagger_{m,n}{\mathcal T_{xy}}\,{\Psi}_{m+1,n+1} \nonumber\\
	&&+ {\psi}^\dagger_{m+1,n}\,{\mathcal{\tilde T}}_{xy}{\psi}_{m,n+1}
	+ \mathrm{H.c.}
	\Big].
\end{eqnarray}
Here
\(
{\psi}_{m,n}=(c^{A\uparrow}_{m,n},c^{B\uparrow}_{m,n},c^{A\downarrow}_{m,n},c^{B\downarrow}_{m,n})^{\mathrm T}
\), where 
$c^{\alpha\beta}_{m,n}$ annihilates a fermion with spin $\beta$ in the $\alpha$ sublattice at site $(m,n)$. The corresponding $4\times4$ matrices are obtained to be
\[
\begin{aligned}
&h_0 = v\,\tau_x\otimes \sigma_0, \qquad
{\mathcal T_x}=(w\,\tau^{-}+\gamma\,\tau^{+})\otimes \sigma_0, \qquad\\[4pt]
&
{\mathcal T_y} = (\delta\,\tau_x + i\tau\,\tau_z)\otimes \sigma_0, \qquad \\[4pt]
&{\mathcal T_{xy}}=-\delta\,\tau^{-}\otimes \sigma_0, \qquad
{\mathcal{\bar T}}_{xy} = -\delta\,\tau^{+}\otimes \sigma_0,
\end{aligned}
\]
with $\tau^\pm=(\tau_x\pm i\tau_y)/2$ and  $\tau_i$' and $\sigma$'s are Pauli matrices in orbital and spin space respectively; $\sigma_0$ and $\tau_0$ are $2\times2$ identity matrix, $v,w$ are the intra and intercell hopping parameters of the standard 1D SSH model. 
Note that the term $i\tau$ breaks time-reversal symmetry (TRS,$\mathcal{T}=i\sigma_y K$). Chiral symmetry ($C=\tau_z\otimes\sigma_z$) survives only when $\tau=0$ and all hopping are real. For $\tau=\delta=0$, we recover the standard 1D SSH model.

To investigate various spin-resolved Chern phases and spin textures of the bulk bands, we now incorporate spin-orbit interaction (SOC).  The tight-binding SOC Hamiltonian written in the same basis as $H_0$ is given by
\begin{align}
H_{\mathrm{SOC}}\label{eq:soc}
&= \sum_{m,n}\Big[
\psi^\dagger_{m,n}\,\Gamma_0\,\psi_{m,n}+\,\psi^\dagger_{m+1,n}\,\Gamma_{xy}\,\psi_{m,n}  \nonumber\\
&+\,\psi^\dagger_{m,n+1}\,\Gamma_{y}\,\psi_{m,n}+\,\psi^\dagger_{m+1,n+1}\,\Gamma_{xy}^{-}\,\psi_{m,n}\nonumber\\
&+\,\psi^\dagger_{m+1,n}\,\Gamma_{xy}^{+}\,\psi_{m,n+1}
\Big].
\end{align}
%(\tau_x\!\otimes\!\Lambda_y)
%(\tau^{+}\!\otimes\!\Lambda_x)
%(\tau_0\!\otimes\!\Lambda_0)
The orbital-spin matrices $\Gamma$s are
	\(\Gamma_0 = \tau_0\!\otimes(-i\alpha\,\sigma_y + i\zeta\,\sigma_z),
		\Gamma_x = \tau_x\!\otimes\,(-i\alpha\,\sigma_y + i\zeta\,\sigma_z),
		\Gamma_y = i\alpha\,\tau_x\,\!\otimes\,\sigma_x,
		\Gamma_{xy}^{-} = i\alpha\,\tau_x\,\!\otimes\ (\sigma^y-\sigma^x),
		\Gamma_{xy}^{+} = i\alpha\,\tau_x\,\!\otimes\ (\sigma^y+\sigma^x).
	\)
	The parameter $\alpha$ represents Rashba-type SOC that mixes spin-up and spin-down states through $\sigma^x$ and $\sigma^y$. In contrast, $\zeta$  correspond to spin-preserving SOC terms along the $x$ direction. Rashba SOC breaks inversion ($\mathcal{P}=i\tau_x\otimes\sigma_z$) and mirror symmetries ($M_x=i\tau_z\otimes\sigma_x$ and $M_y=\tau_y\otimes\sigma_y$), and they remove any spin conservation. Note that the terms $\Gamma_y$, $\Gamma_{xy}^{-}$, and $\Gamma_{xy}^{+}$ give rise to spin-dependent inter-chain couplings, which control the evolution of spin textures between adjacent SSH chains.
    
Together with Eq.~(\ref{eqn:main_ham}) and (\ref{eq:soc}), the full Hamiltonian in momentum space becomes
	\begin{align}\label{eq:momentumHam}
		H &= \sum_{\mathbf{k}}
		\Big[
		2\tau\sin k_y\,\tau_z\mathbb{I}_2
		+ \tau^+M(\mathbf{k})
		+ \tau^- M^\dagger(\mathbf{k})
		\Big],
	\end{align}   
	where $\mathbf{k}=(k_x,k_y)$ and the $2\times2$ spin-space coupling matrix is
	\[
	M(\mathbf{k})=
	A(\mathbf{k})\,\mathbb{I}_2
	+ B(\mathbf{k})\,\sigma_z
	+ D_x(\mathbf{k})\,\sigma_x
	+ D_y(\mathbf{k})\,\sigma_y,
	\]
	with
	\(		A(\mathbf{k}) = v
		+ w\,e^{-ik_x}
		+ \gamma\,e^{ik_x}
		+ 2 \delta\cos k_y \,(1-e^{-ik_x}),
		B(\mathbf{k}) = i\!\left(\zeta - \zeta e^{-ik_x}\right),
        D_x(\mathbf{k}) = 2\alpha\!\left(\sin k_y 
		+ i\cos k_y\, e^{-ik_x}\right),
		D_y(\mathbf{k}) = -i\alpha\,(1-e^{-ik_x}) + 2\alpha e^{-ik_x}\sin k_y.
		\)

Introducing complex hopping {\it only} along the SSH chains replaces $\omega\rightarrow\omega+i\eta_1$ and $\gamma\rightarrow\gamma+i\eta_2$ for forward hopping. Note that, addition of $\eta_1$ and $\eta_2$ along with SOC leads to an extended phase diagram with different topological number as will be evident shortly. Note also, complex hopping amplitudes break inversion and mirror symmetries of the lattice construction. Additionally, they break the TRS symmetry even with $\tau=0$.  Thus, for all parameters with finite values, the model belongs to symmetry class~A  with no TRS, inversion, or chiral symmetries. We shall show that only a special regime of these unconventional topology can host persistent spin-texture in the presence of finite RSOC. Thus the present setting provide us a platform to explore persistent spin-texture in a topological background as compared to conventional 2D semiconductors.

%\end{comment}
\section{Distinct Topological Phases }
Let us first chart out distinct topological phases for both in the absence and in the presence of different SOC and complex hopping terms. Wherever possible, we derive effective low energy model Hamiltonians for different choices of parameters and  express Chern numbers explicitly in terms of model parameters. 
\subsection{Without RSOC}
\subsubsection{Clean limit}
In the absence of SOC ($\zeta=0, \alpha=0$), second nearest neighbour and complex ($\gamma=0,\,\eta_1=0,\,\,\eta_2=0$) hopping, the  Hamiltonian in Eq.~(\ref{eq:momentumHam}) is block diagonal and can be expressed as 
\begin{align}
H(\bm{k}) = h_{x}(\bm{k})\,\tau_x + h_{y}(\bm{k})\,\tau_y + h_z(\bm{k})\,\tau_z,
\end{align}
where
\begin{align}
&h_x = v + 2\delta \cos k_y + (w-2\delta \cos k_y)\cos k_x,\nonumber\\
&h_y = (w-2\delta \cos k_y)\sin k_x,\nonumber\\
&h_z = 2\tau \sin k_y.
\end{align}

A massive Dirac point $\bm{k}_D$ is defined by the simultaneous conditions of $h_x(\bm{k}_D)=h_y(\bm{k}_D)=0$ with the non-zero mass term $\sin k_{Dy}\neq0$. A small deviation from the Dirac point, i. e., $\bm{q}=\bm{k}-\bm{k}_D$  allows to derive the low-energy theory for various cases, which in term leads to find phase boundary of topological phases. 

\begin{figure}
\includegraphics[width=1.0\linewidth]{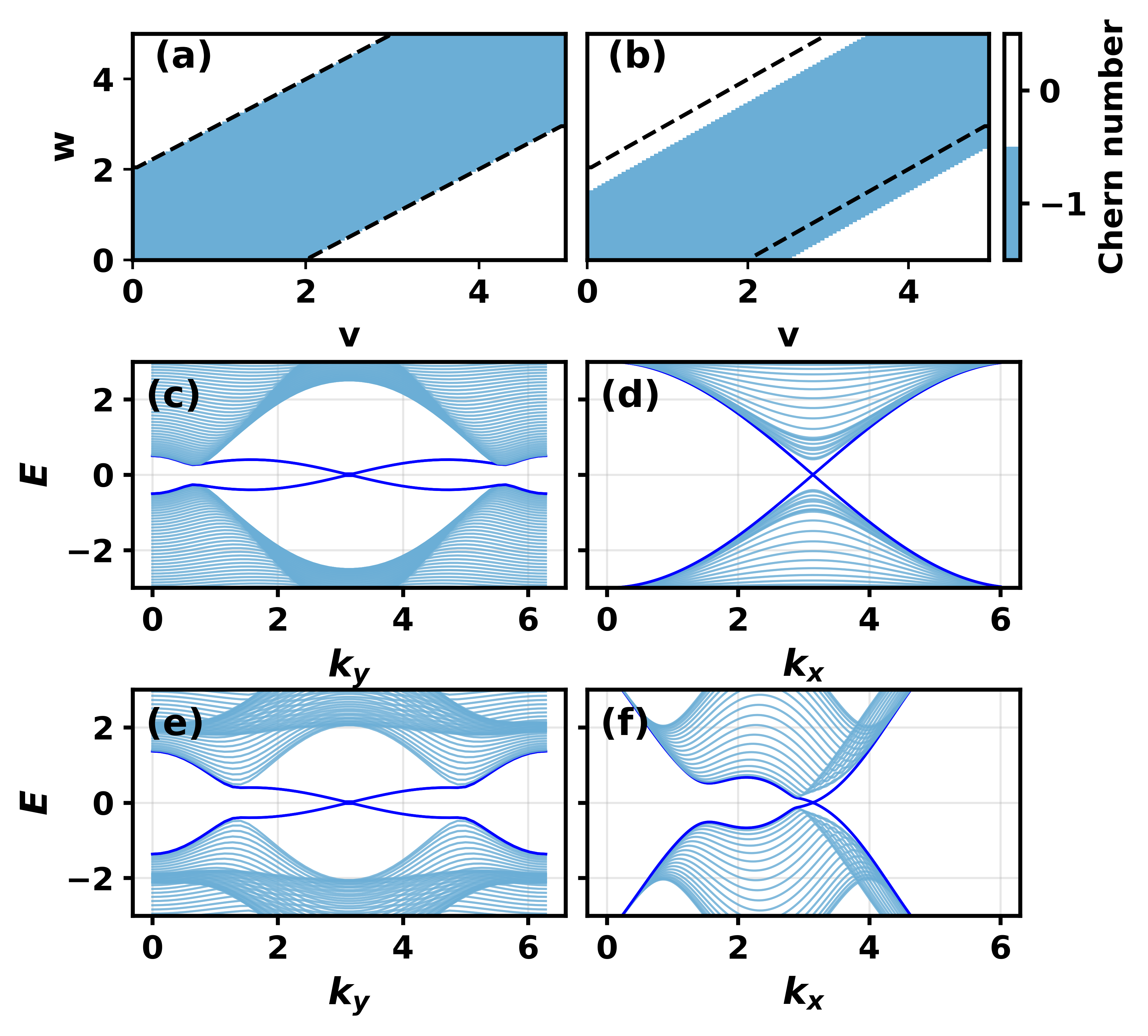}
\caption{a) The topological phase diagram from the full lattice Hamiltonian (Eq.~\ref{eqn:main_ham}) in the plane of model parameters $v$ and $w$ of the standard 1D SSH. Here we set $\delta = 0.5,\tau = 0.2$. The black solid line represent the phase boundary obtained from the expression of Dirac points $k_{D_y}$.b) Same as (a) once we introduce long-range hopping $\gamma=0.5,\eta_{1,2}=0$. Fig (c-f) shows dispersion relation with $w = 2, v = 1 $. For Fig (c,d) $\gamma = 0, \eta = 0$ , Fig (e,f) $\gamma = 0.5, \eta = 1$  }
\label{fig2:Chern1andspec}
\end{figure}

For $v-4\delta\leq w\leq v+4\delta$, the conditions for Dirac points yield $k_{Dx}=\pi$ and $ k_{Dy}=\pm\cos^{-1}((w-v)/4\delta)$ with $\sin k_{Dy}\neq0$. Expanding the Hamiltonian around $\bm{k}_D$ up to leading order, we obtain

\begin{align}
\label{eq:lowenergy}
H_{\mathrm{Dirac}}^{(I)}(\bm{q})
=v_x\, q_y\, \tau_x+v_y
\, q_x\,\tau_y
+m\, \tau_z.
\end{align}
where $v_x=-4\delta \sin k_{Dy}$, $v_y=-(w 
+v)/2$ and $m=2\tau\,\sin k_{D_y}$. Clearly,  the velocities of Dirac quasiparticles are asymmetric in $k_x$ and $k_y$, which in turn, leads to an asymmetric spectrum (see Fig.~(\ref{fig2:Chern1andspec})). 

To find topological phases of the effective low-energy Hamiltonian $H_{\mathrm{Dirac}}^{(I)}(\bm{q})$, we define Chern number ($C$) as
\begin{align}
C =\frac{1}{2}\sum_{\bm{k}_i \in k_D} \sgn\!\left[
		\left(
		\partial_{k_x}\bm{h}
		\times
		\partial_{k_y}\bm{h}
		\right)_z
		\right]_{\bm{k}_i}
		\;
		\sgn\!\left[
		h_z(\bm{k}_i)
		\right].
	\label{eq:Chern_Dirac}
\end{align}
The first term \(
\sgn\!\big[
(\partial_{k_x}\bm{h}\times\partial_{k_y}\bm{h})_z
\big]
\)
encodes the chirality of the Dirac point and is determined by the Jacobian of the mapping from momentum space to the Bloch sphere, while the second term
\(
\sgn[h_z(\bm{k}_i)],
\)
corresponds to the sign of the Dirac mass evaluated at the Dirac point. The total Chern number is obtained by summing over contributions from all Dirac points in the Brillouin zone. For Eq.~(\ref{eq:lowenergy}), notice that the Dirac velocity $v_x$ and the Dirac mass $m$ changes sign between two Dirac points $\pm k_{D_y}$. Thus, both the Dirac points contribute $-0.5$ to  $C$ and the total Chern number is obtained to be $-1$. This applies only to the regime for which $v-4\delta\leq w\leq v+4\delta$ for each of the spins. The equality determines the {\it phase boundary} across which the topological transitions occur. 

Fig.~(\ref{fig2:Chern1andspec})a, shows topological phase diagram obtained numerically in the $w-v$ plane for fixed $\delta$ and $\tau$ in the clean limit. Evidently, we obtain topological phase with Chern number $C=-1$ for each of the spin, leading to {\it standard} quantum anomalous Hall phases, as denoted by $(C_{\uparrow}=-1,C_{\downarrow}=-1)$. To further complement the presence of Chern phases, we diagonalize the Hamiltonian in Eq.~\ref{eqn:main_ham} with open boundary along $x$ and $y$ separately, as the bulk spectrum is anisotropic. Fig.~(\ref{fig2:Chern1andspec}) (c-d) evidences the anisotropic spectrum along $k_x$ and $k_y$ and the presence of chiral edge modes in the clean limit. 

\subsubsection{$\gamma\ne0$ and $\eta_1\ne0,\eta_2\ne 0$, $\zeta=0$}
When higher-order hopping term $\gamma$, together with the
imaginary components $\eta_1,\eta_2$, is incorporated, the Dirac point $\bm{k}_D$ remains located at $k_{Dx}=\pi$ but is shifted along $k_y$ direction, satisfying $\cos k_{Dy}=(w+\gamma-v)/(4\delta)$. Expanding about $k_D$, the low energy Hamiltonian up to the leading order is found to be 
\begin{align}
&H_{\mathrm{Dirac}}^{(II)}(\bm{q})\nonumber\\
&=H_{\mathrm{Dirac}}^{(I)}(\bm{q})-(\eta_1+\eta_2)\, q_x\,\tau_x-(\eta_1-\eta_2+\gamma\, q_x)\, \tau_y.
\end{align}
We note that the finite $\eta_1,\eta_2$ generate an additional $\tau_x$ term linear in $q_x$, thereby mixing the momentum $q_x$ and $q_y$. Moreover, when $\eta_2\ne\eta_1$, the leading-order contribution to $\tau_y$ becomes momentum independent, which in turns gives rise to unusual spin-texture, as will be evident once we introduce Rashba SOC ($\alpha\ne0$). Overall, finite $\eta_2$, $\eta_1$ and $\gamma$ merely renormalizes the effective velocities without opening a gap.
Consequently, their presence does not alter the topology of the system, apart from shifting the phase boundaries. This shift of the Chern phases, without the emergence of any additional topological phase, is clearly illustrated in Fig.~\ref{fig2:Chern1andspec}(b). Accordingly, the shift in the band minima along $k_x$ is evident from Fig.~\ref{fig2:Chern1andspec}(e-f).

\subsubsection{($\zeta,\eta_1,\eta_2,\gamma)\ne 0$}
\begin{figure}
\includegraphics[width=1.0\linewidth]{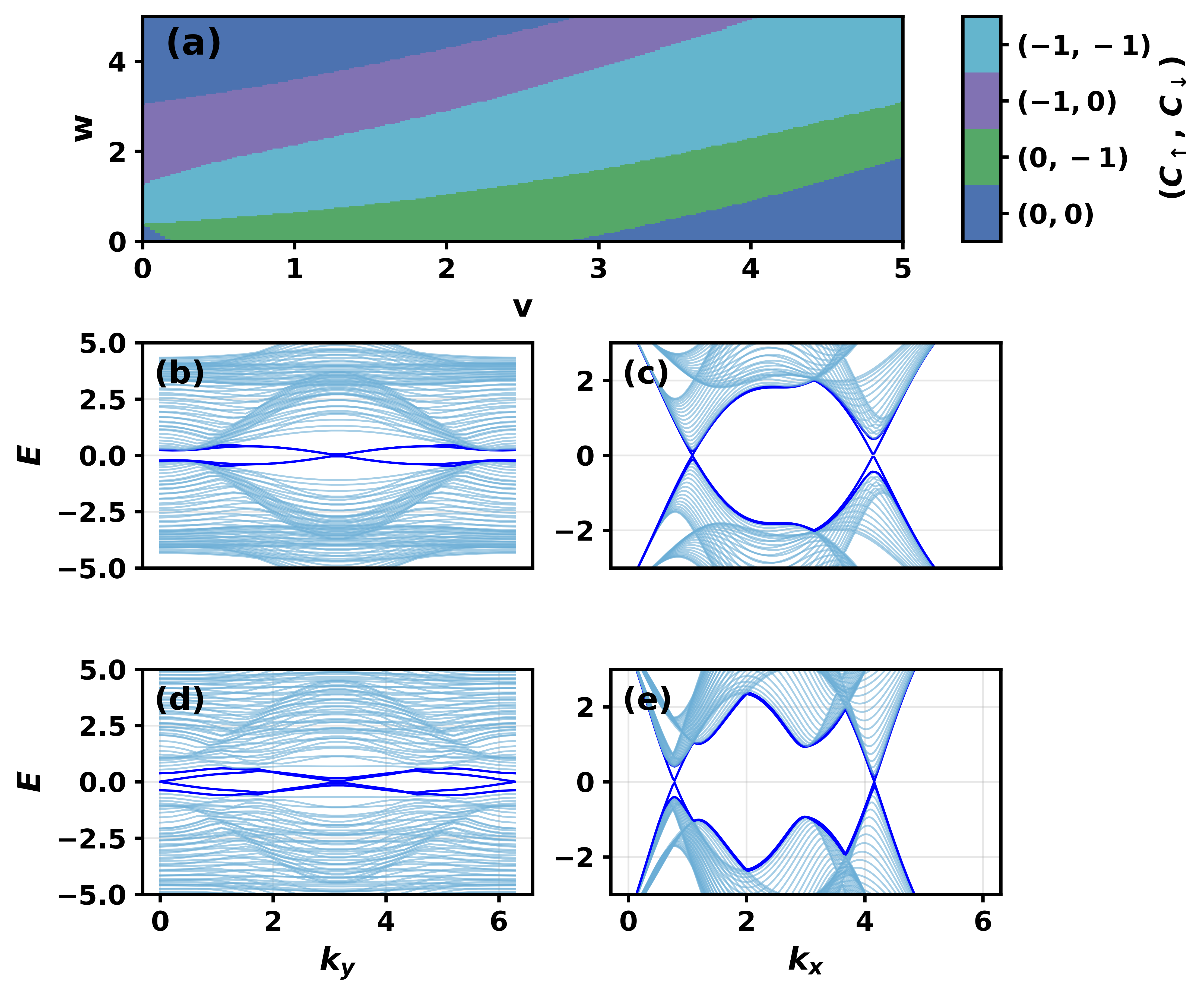}
\caption{ The topological phase diagram from the full lattice Hamiltonian in the plane of model parameters $v$ and $w$ standard 1D SSH with spin conserving SOI term $\zeta = 1$ . Here we have $\delta = 0.5, \tau = 0.2, \gamma = 0.5$.Fig (c-f) shows dispersion relation with $w = 2, v = 1 $. For Fig (a,c,d) $\eta_1=  \eta_2 = 1$ , Fig (b,e,f) $\eta_1 = 1, \eta_2 = 2$  }
\label{fig3:ChernANDspectrum}
\end{figure}

We now turn to find the effective low-energy Dirac Hamiltonian in the presence of the spin conserving SOC, i. e., $\zeta\ne 0$. This introduces an additional term $-s\zeta\,\sin k_x$ to $h_x$ and $-s\zeta\,(1-\cos k_x)$ to $h_y$, while the full Hamiltonian remains spin($s=\pm$)-resolved. This shifts the location of Dirac points along $k_x$ and $k_y$ from the previous cases and complicates the simple equations for $k_{D_x}, k_{D_y}$, which we do not present here for simplicity. However, linearising the Hamiltonian for finite $\gamma,\eta_1,\eta_2,\zeta$ around the Dirac point $k_D$ for the spin $s$, we obtain
\begin{align}
&h_{x,s} \simeq A_{x,s} q_x + B_x q_y\nonumber\\
&h_y \simeq A_{y,s} q_x,
\label{eq:lowSOC}
\end{align}
where \(
A_{x,s} = -(w+\gamma-2\delta \cos k_{Dy})\sin k_{Dx} + (\eta_1+\eta_2+s\zeta)\cos k_{Dx},
\)
\(
B_x = -2\delta \sin k_{Dy}(\cos k_{Dx}-1),
\)
\(
A_{y,s} = (w-\gamma-2\delta \cos k_{Dy})\cos k_{Dx} + (\eta_2-\eta_1)\cos k_{D_x}+s\zeta \sin k_{Dx},
\)
\textcolor{red}{.} Clearly, the low-energy effective Dirac Hamiltonian $h_x$ in Eq.~(\ref{eq:lowSOC}) mixes $q_x$ and $q_y$ as before. Interestingly, the presence of $\eta_1$, $\eta_2$ and $\zeta$ together {\it break} the spin degeneracy of the effective Hamiltonian. This in turn leads to asymmetric Chern phase diagram as evident in Fig.~\ref{fig3:ChernANDspectrum}(a). Correspondingly, for specific $v$  and $w$, we find $(C_{\uparrow}=-1,0)$ and $(0,C_{\downarrow}=-1)$ phases as marked by green and blue regions in Fig.~\ref{fig3:ChernANDspectrum}(a). We note that this is one of the main results of this study. This is typically termed as {\it quantum spin anomalous Hall insulating} (QASHI) phases and can give rise to spin-filtered spin transport. This clarifies the interesting interplay between complex hopping and spin-orbit coupling as pointed out in the introduction. 

Figures~\ref{fig3:ChernANDspectrum}(b-c) show the bulk and edge energy spectra for $\eta=\eta_1=\eta_2=1$ and $\zeta=1$, illustrating how the presence of complex hopping and spin-orbit coupling modifies the band structure. Figure~\ref{fig3:ChernANDspectrum}(b) shows the energy spectrum as a function of $k_y$ with open boundary conditions along the $x$ direction. The bulk bands display a clear anisotropic energy gap, with in-gap edge-localized states between lowest conduction and highest valence bands. Note that, the in-gap states are associated with only one spin sector with a nonzero topological invariant, while the other spin sector remains topologically trivial and does not support edge modes. This spin-selective behavior originates only when both $\eta$ and $\zeta$ are finite. 

Figure~\ref{fig3:ChernANDspectrum}(c) on the other hand presents the energy spectrum obtained by imposing open boundary along the $y$ direction. Compared to  Fig~\ref{fig3:ChernANDspectrum}(b), the changes in the energy bands are more pronounced along the $k_x$ direction. This is a manifestation of stronger momentum dependence along $k_x$ due to the presence of $\eta$ and $\zeta$ as can be seen in Eq.~(\ref{eq:lowSOC}).

Figure~\ref{fig3:ChernANDspectrum}(d-e) correspond to the case where $\eta_1 \neq \eta_2$. In this regime, the asymmetry between the imaginary hopping amplitudes introduces additional contribution to $A_y,s$ in Eq.~(\ref{eq:lowSOC}). Consequently, the bulk energy spectra for both $k_y$ and $k_x$ further distorted compared to Fig.~\ref{fig3:ChernANDspectrum} (b-c).

\begin{figure}
\includegraphics[width=\linewidth]{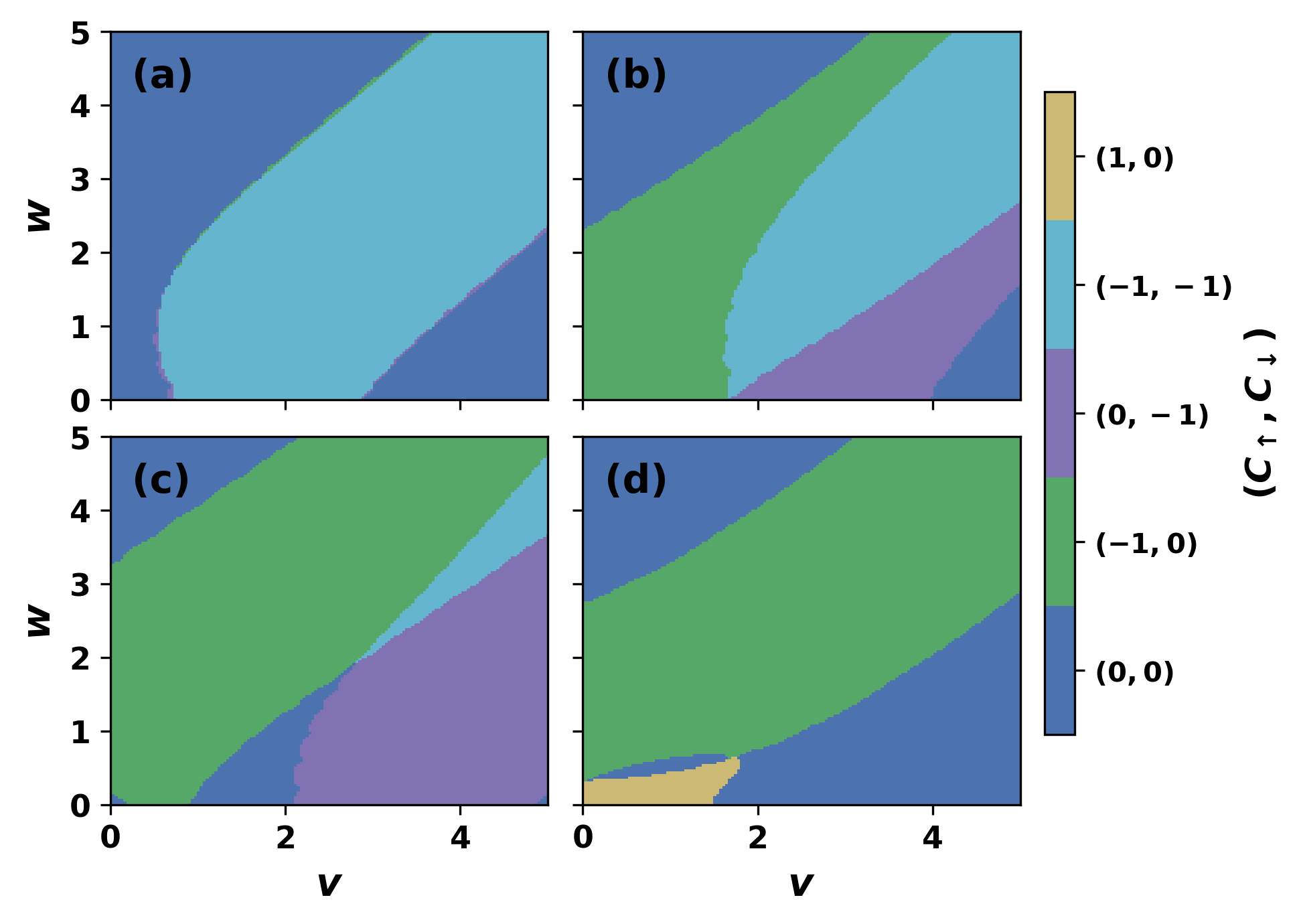}

\caption{The topological phase diagram from the full lattice Hamiltonian in  $v-w$ plane for a set of four parameters $(\eta_1,\eta_2)=(0,0)$ (a), $(1,0)$ (b), $(1,1)$ (c) and $(4,1)$ (d) with spin conserving SOI term $\zeta = 1$ and RSOI term $\alpha = 1$  . Here we have $\delta = 0.5, \tau = 0.2, \gamma = 0.5$.}
\label{fig:Chernforvariouscomplexhop}
\end{figure}

\subsection{With RSOC}
\subsubsection{$\zeta,\eta,\gamma\ne 0$,$\alpha\ne0$}
Compared to the case with $\alpha=0$, the inclusion of the Rashba term $H_{\mathrm{SOC}}$ introduces spin-flip processes through $D_x(\bf{k})$ and $D_y(\bf{k})$ in Eq.~(\ref{eq:momentumHam}), thereby mixing the $\uparrow$ and $\downarrow$ spin sectors. As a consequence, the spin Chern number discussed above looses in meaning as spin in no longer conserved. Nevertheless, the underlying topological properties of the system can still be characterized using an effective spin Chern number, evaluated numerically using a projection-based approach\cite{prodan-2009,prodan-2010}. We first construct the projector onto the occupied subspace, \begin{equation} P(\bm{k}) = \ket{n_1(\bm{k})}\bra{n_1(\bm{k})} + \ket{n_2(\bm{k})}\bra{n_2(\bm{k})}, \end{equation} where $\ket{n_{1,2}(\bm{k})}$ are the Bloch states of the Hamiltonian in Eq.~(\ref{eq:momentumHam}), corresponding to the two valence bands with energies $E_{1,2}(\bm{k})<0$. Using $P(\bm{ k})$, we define the projected spin operator \begin{equation} \tilde{S}(\bm{k}) = P(\bm{k}) \, (\tau_3 \otimes \sigma_0) \, P(\bm{k}), \end{equation} which acts within the four-dimensional occupied subspace. Diagonalization of $\tilde{S}(\bm{k})$ yields four eigenstates $\ket{\psi_n(\bm{k})}$ with eigenvalues $\epsilon_n(\bm{k})$ $(n=1,\ldots,4)$, where $|\epsilon_1|=|\epsilon_4|\neq0$ and $|\epsilon_2|=|\epsilon_3|\simeq0$ within numerical accuracy, with $\epsilon_1(\bm{k})<0$ and $\epsilon_4(\bm{k})>0$. Together with the eigenstates $\ket{\psi_1(\bm{k})}$, $\ket{\psi_4(\bm{k})}$ associated with the nonzero eigenvalues and gauge-invariant lattice formulation in $\bm{k}$-space proposed by Fukui \emph{et al.}~\cite{spin-chern2}, we numerically compute the effective spin-resolved Chern numbers $C_\uparrow$ and $C_\downarrow$.

\begin{figure}
\includegraphics[width=1.0\linewidth]{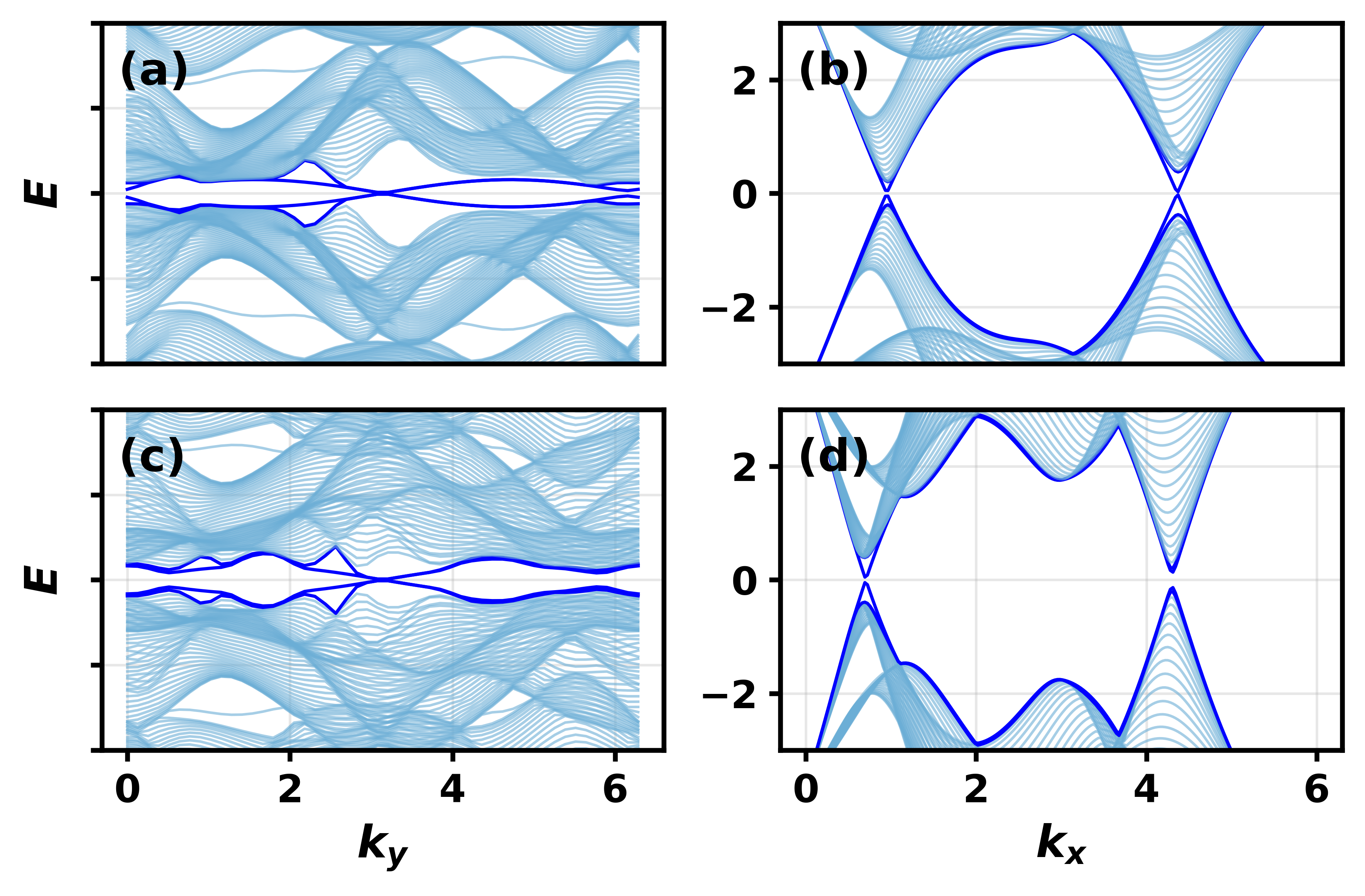}
\caption{Energy spectrum for spin conserving SOI term $\zeta = 1$ and RSOI term $\alpha = 1$ . Here we have $\delta = 0.5, \tau = 0.2, \gamma = 0.5$. Fig (c-f) shows dispersion relation with $w = 2, v = 1 $. For Fig (c,d) $\eta_1= \eta_2 = 1$ , Fig (e,f) $\eta_1 = 1, \eta_2 = 2$,$\alpha = 1$ }
\label{fig4:ChernANDSpectrum}
\end{figure}

Figure~\ref{fig:Chernforvariouscomplexhop} illustrates topological phase diagram in the presence of spin--orbit interaction (RSOC) and various values of complex hopping $\eta_1,\eta_2$. The impact of $\alpha$ and $\eta_1,\eta_2$ is evident in the phase diagram. For $\eta_1=\eta_2=0$, we obtain standard QAHI phase with $C_{\uparrow}=-1,C_{\downarrow}=-1$, corroborating the standard cases discussed before. In contrast,  for finite $\eta_1, \eta_2$, we obtain all three distinct phases (see Fig.~\ref{fig:Chernforvariouscomplexhop}b-d) similar to case 3 discussed in the preceding section; however phase boundaries are modifies depending on the relative strength of $\eta_1$ and $\eta_2$ in the same parameter regime of $v-w$. Notice that for $\eta_1\gg \eta_2$, there is a change in the {\it chirality} as we obtain $(C_{\uparrow}=1,C_{\downarrow}=0)$ instead of $(C_{\uparrow}=-1,C_{\downarrow}=0)$ for small values of $u-v$ (Fig.~\ref{fig:Chernforvariouscomplexhop}d). In sum, the phase space for both the $(C_{\uparrow}=1,0)$ and $(0,C_{\downarrow}=1)$ can be enhanced and tuned by the RSOC and complex hopping as opposed to the $\alpha=0$ case. 

Figures~\ref{fig4:ChernANDSpectrum}(a–d) present the energy spectrum for $\alpha \neq 0$ for two representative choices of complex hopping parameters: $(\eta_1=\eta_2=1)$ and $(\eta_1=1,\eta_2=2)$. As seen in Figs.~\ref{fig4:ChernANDSpectrum}(a) and (c), the dispersion as a function of $k_y$ is markedly modified compared to the $\alpha=0$ case, exhibiting substantial restructuring of the bulk bands. Notably, edge-like branches (indicated by arrows) emerge within the bulk gap, signaling the presence of Rashba-SOC driven nontrivial floating states. In contrast, the energy dispersions as a function of $k_x$, shown in Figs.~\ref{fig4:ChernANDSpectrum}(b) and (d), exhibit qualitatively similar features to those observed for $\alpha=0$, with no significant reconstruction of the band structure. This anisotropic feature highlights the distinct roles played by $\alpha$ and the complex hopping parameters in shaping the spectral and topological properties of the system.

%{\color{red}The effect of Rashba spin--orbit interaction is also evident in the energy spectra shown in Figs.~\ref{fig4:ChernANDSpectrum}(b--e). The band crossings present in the $\alpha=0$ case are partially lifted, leading to avoided crossings in the low-energy spectrum, and edge states that were spin selective acquire mixed spin character. Although the bulk energy gap remains open, the dispersion and overall structure of the low-energy bands are noticeably altered.
%Figures~\ref{fig4:ChernANDSpectrum}(b) and (d) show the energy spectra as a function of $k_y$ with open boundary along the $x$ direction. While the overall envelope of the band structure and the bulk energy gap remain similar to $\eta_1=\eta_2$ and $\eta_1\neq\eta_2$, additional Rashba-split bulk bands appears which exhibit  edge-like branches (indicated by arrow) within the bulk spectrum. 
%Figs.~\ref{fig4:ChernANDSpectrum}(c) and (e) display the energy dispersion as a function of $k_x$, exhibit more pronounced modifications similar to the $\alpha=0$ case.} 

\begin{figure}
\includegraphics[width=\linewidth]{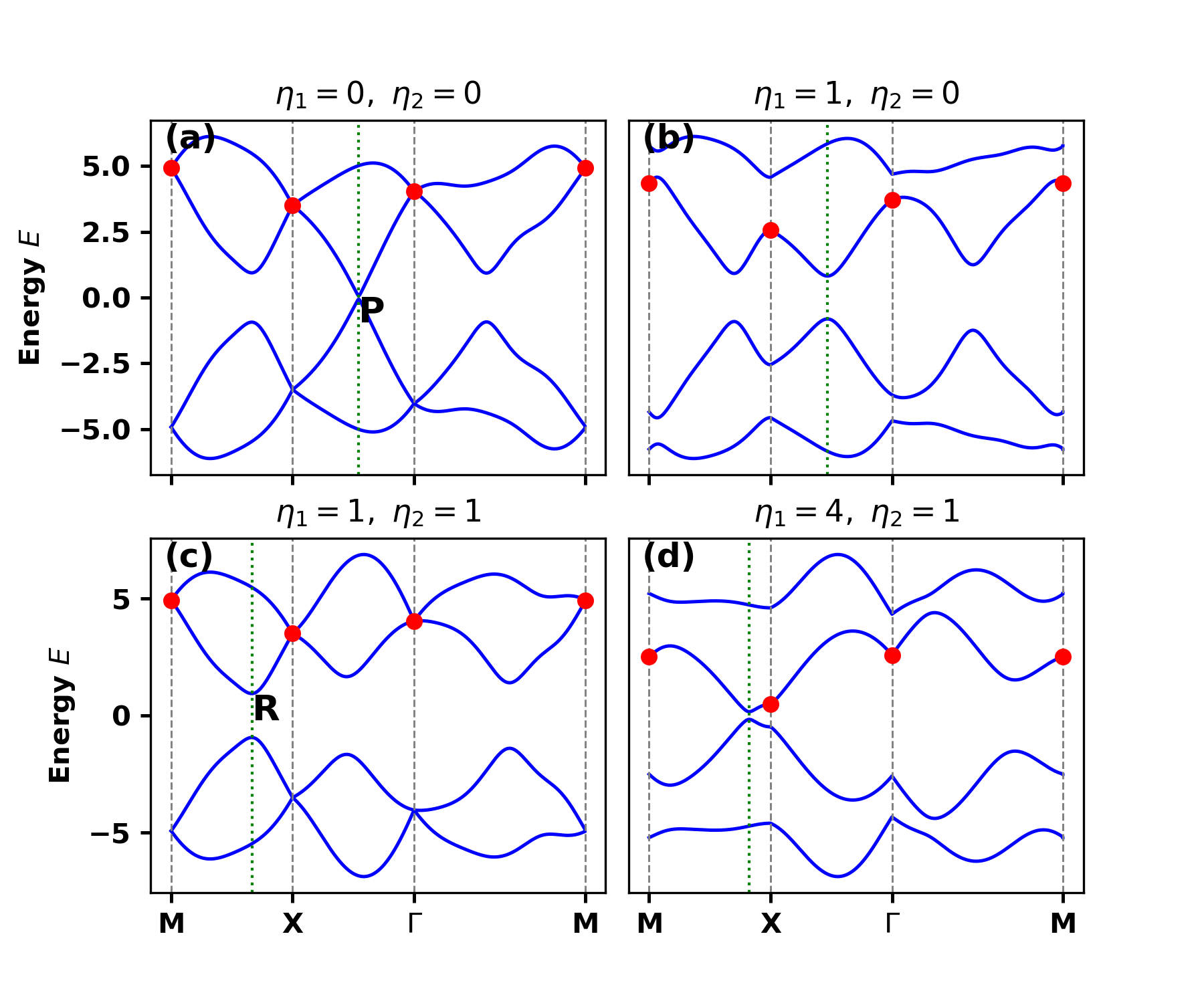}
\caption{ Band spectrum calculated along high-symmetry lines for four different sets of (a) $\eta_1=0,\eta_2=0$, (b)$\eta_1=1,\eta_2=0$, (c) $\eta_1=1,\eta_2=1$, and (d) $\eta_1=4,\eta_2=1$. The dashed vertical lines are for eye guidance for the high symmetry $k-$ points. For $\eta_1=\eta_2$, both conduction and valence bands are degenerate at the high symmetry points. Here the P and R  gives the minimum of conduction for $\eta = 0,1$ band. }
\label{fig4:spectrumHighsymmetry}
\end{figure}
\section{Spin Texture}
We next investigate if the topological phases discussed above can host persistent spin texture\cite{note1}, despite all symmetries are absent due to the presence complex hopping and spin-orbit coupling pointed out before. To this end, we focus on the spectrum at the high-symmetry ${\bf k}$-points in the Brillouin zone. This is because PSTs are typically known to emerge in the vicinity of such points in the BZ\cite{li_APL2024,Autiteri_PRM,djani_QM2019,Zhang2022,PhysRevB.110.235162,Jin_Nano2021,slawniska_2Dmat}. Deviations from this conventional behavior have also been reported, for instance in WTe$_2$\cite{jose_prl2020}, where a canted persistent spin structure appears at a finite momentum $Q=(0.233,0)$. In this spirit, Fig.~\ref{fig4:spectrumHighsymmetry} shows the energy spectrum along the high symmetry directions ($M\rightarrow X\rightarrow\Gamma\rightarrow M$) for four representative choices of ($\eta_1,\eta_2$).
Interestingly, for $\eta_1=\eta_2$, both the conduction and valence bands at the high-symmetry points are found to be degenerate, reflecting the presence of time-reversal symmetry (TRS) at these points. Moreover, the minimum of the conduction band occurs at a finite momentum $P=(1.44,0)$ $R=(3.14, 1.05)$ rather than at any high-symmetry $\bm k$-points, as shown in Fig.~\ref{fig4:spectrumHighsymmetry}a (Fig.~\ref{fig4:spectrumHighsymmetry}c). For $\eta_1\neq\eta_2$, these degeneracies are lifted, although the band minima remain located away from the high-symmetry points. In contrast, when $\eta_1\gg\eta_2$, the band minimum appears predominantly around the $X$ and $\Gamma$ points. In the next, we analyze the spin texture both near high-symmetry $\bm{k}$ points and around band minima located away from them. 
%\subsubsection{$\eta_1=\eta_2$}

\begin{figure*}
\includegraphics[width=\linewidth]{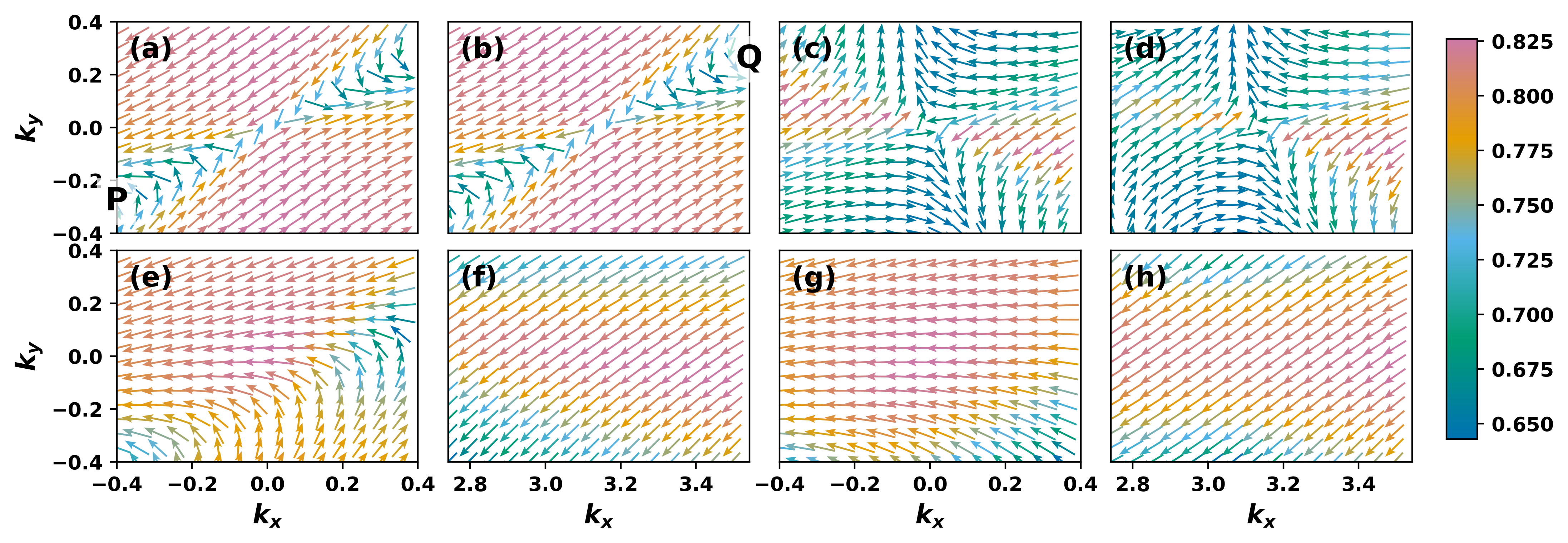}
\caption{(a,b)The spin structures around $\Gamma$ for $\eta=0$ and $\eta=1$, respectively. (c,d) Same as (a,b) around $X$. For $\eta_1=1>\eta_2=0$, the quasi-persistent nature of the spin textures around $\Gamma$ (e) and $X$ (f) is evident. In contrast, for $\eta_1=4\gg \eta_2=1$, the relatively stronger unidirectional spin structures are evident around $\Gamma$ (g) and $X$ (h). All other parameters are taken same as Fig.~\ref{fig:Chernforvariouscomplexhop}.}
\label{fig:spintextureGamma}
\end{figure*}

Let us first discuss the case for $\eta=\eta_1=\eta_2$. In this regime, all high-symmetry points are twofold degenerate, and the lowest conduction band attains its maximum at these points. The in-plane spin texture, defined as
\begin{align}\vec\sigma=(\langle\psi|\tau_0\otimes\sigma_x|\psi\rangle,\langle\psi|\tau_0\otimes\sigma_y|\psi\rangle),
\end{align} where $|\psi\rangle$ denotes the eigenstate of the lowest conduction band, is found to wind near the high-symmetry points. Figure~\ref{fig:spintextureGamma} (a-b) illustrates the spin textures around the $\Gamma$ for $\eta=0$ and $\eta=1$, respectively, while Fig.~\ref{fig:spintextureGamma} (c-d) represents the same around $X$ points. Clearly, in all cases, the spin textures exhibit finite rotations in the $x$–$y$ plane. The origin of this behavior is attributed to $\eta$ independent effective Hamiltonian in spin space, which will be derived and discussed in the following sections.

%\subsection{$\eta_1\ne\eta_2$}
For $\eta_1\neq\eta_2$ with $\eta_1>\eta_2$, the twofold band degeneracies are lifted around the $X$, $\Gamma$, and $M$ points. Nevertheless, the lowest conduction band still exhibits maxima in the vicinity of these points. Despite the presence of band maxima, we present corresponding spin textures in Fig.~\ref{fig:spintextureGamma}(e,f) for $\eta_1=1$ and $\eta_2=0$ for completeness. Interestingly, the spin textures around the $\Gamma$ and $X$ points display a quasi-persistent character, where the spin expectation values vary significantly in magnitude but undergo only weak rotations in the $x$–$y$ plane.

In contrast, for $\eta_1\neq\eta_2$ with $\eta_1\gg\eta_2$, the band minimum is found to occur near the $\Gamma$ and $X$ points. Remarkably, the persistent nature of the spin structures as shown in Fig.~\ref{fig:spintextureGamma}(g,h) become more pronounced in the vicinity of this band minimum. Similar to earlier cases, the origin of this behavior can be elucidated from the approximate effective Hamiltonian, which will be discussed in the following sections.

Finally, before moving on to the next sections, we note that the spin texture near the $M$ point behaves similarly to that near the $X$ point and is therefore not shown explicitly. In contrast, the spin textures around the points $P$ and $R$ remain partially unidirectional but the spin magnitude differs significantly as shown in Fig.~\ref{fig:spintexturePQ}(a,b). This is in contrast to cases for same values of $\eta$, i. e., $\eta=0$ and $\eta=1$ at the high-symmetry points.

\subsection {Derivation of Effective Hamiltonian}
To gain insight into why $\eta_2\neq\eta_1$ leads to unconventional spin configurations around the $X$, $\Gamma$, and $M$ points, we derive an effective spin Hamiltonian perturbatively from the four-band spin–orbit–coupled Hamiltonian. To this end, we rewrite the Hamiltonian in Eq.~(\ref{eqn:main_ham}) in the spin $\otimes$ orbital basis and decompose the Hilbert space as a direct sum of two subspaces labeled by $\alpha$ and $\beta$, where $\alpha$ and $\beta$ denote the spin and orbital degrees of freedom, respectively. The Hamiltonian then takes the following form:

\begin{align}
H=\begin{pmatrix}H_{\alpha\alpha}& H_{\alpha\beta}\cr
H_{\beta\alpha}&H_{\beta\beta}
\end{pmatrix},
\end{align}
where 
\begin{align}\label{eq:pertubative}
&H_{\alpha\alpha}=\begin{pmatrix}2\tau \sin k_y &0\cr 0&2\tau\sin k_y\end{pmatrix},\nonumber\\&H_{\alpha\beta}=\begin{pmatrix}A(k)+B(k)& D_x(k)-i D_y(k)\cr D_x(k)+i D_y(k) &A(k)-B(k)\end{pmatrix}
\end{align}
with $H_{\beta\beta}=-H_{\alpha\alpha}$ and $H_{\beta\alpha}$ is same as $H_{\alpha\beta}$ with all elements are replaced by their complex conjugate. 
Following Ref.\onlinecite{lowdin}, it is easy to show that the eigenvalue equation $H|\psi\rangle=E|\psi\rangle$ leads to an effective Hamiltonian in the $\alpha$ space as 
\begin{align}
H_{\alpha}=H_{\alpha\alpha}+H_{\alpha\beta} (E-H_{\beta\beta})^{-1}H_{\beta\alpha}
\end{align}
with $|\psi\rangle$ having the form $|\psi\rangle=(|\psi_{\alpha}\rangle,|\psi_{\beta}\rangle)^T$ and $E$ is the relevant energy for the effective Hamiltonian. Using Eq.~\ref{eq:pertubative}, a straight forward calculation (see Appendix \ref{app:dirac_derivation}) leads to an effective Hamiltonian in spin space as 
\begin{align}
H_{\rm eff}\simeq d_x\sigma_x+d_y\sigma_y+d_z\sigma_z
\label{eq:eff}
\end{align}
where
\begin{align}
&d_x=\frac{2}{E+2\tau\sin k_y}[{\rm Re}(A(k)D_x(k)^{\ast})-  {\rm Im}(B(k)^{\ast}D_y(k))]\nonumber\\
&d_y=\frac{2}{E+2\tau\sin k_y}[{\rm Re}(A(k)D_y(k)^{\ast})- {\rm Im}(B(k)D_x(k)^{\ast})]\nonumber\\
&d_z=\frac{2}{E+2\tau\sin k_y}[{\rm Re}(A(k)B(k)^{\ast})-{\rm Im} (D_x(k)D_y(k)^{\ast})],
\end{align}
where $\rm Re$ and $\rm Im$ refer to real and imaginary part of a complex number. Note that we have neglected terms that appear with identity matrix $\sigma_0$ as it does not affect the eigenstates.

\subsubsection {Approximate effective Hamiltonian around $\Gamma=(0,0)$}
Having derived the generic $2\times2$ Hamiltonian in spin space, we next obtain approximate effective Hamiltonians around the $\Gamma$ and $X$ points and briefly comment on the $M$, $P$, and $Q$ points. Linearizing Eq.~(\ref{eq:eff}) about $\Gamma=(0,0)$, the effective Hamiltonian takes the form
\begin{align}
H_{\rm low}^{\Gamma}(q_x,q_y)
&=(f_1' q_x+f_2' q_y)\sigma_y
+(f_0' q_y +f_3 q_x)\sigma_z \nonumber\\
&\quad+
\big[
f_0(\eta_1-\eta_2)
+f_1 q_x
+f_2 q_y
\big]\sigma_x ,
\end{align}
where $f,f_0,f_1,f_2,f_1',f_2',f_3$ are functions of the parameters $v,w,\gamma,\delta,\tau,\alpha$, and $\zeta$ (see Appendix \ref{app:Gamma}). 

For $\eta_1=\eta_2$, the leading-order contributions to all Pauli matrices are momentum dependent, resulting in a strongly momentum-dependent spin texture around the $\Gamma$ point, as shown in Fig.~\ref{fig:spintextureGamma}(a,b). In contrast, for $\eta_1\neq\eta_2$ with $\eta_1\gg\eta_2$, the momentum dependence in the coefficient of $\sigma_x$ can be neglected to leading order. Nevertheless, the remaining momentum-dependent terms associated with $\sigma_y$ and $\sigma_z$ act together to suppress strictly unidirectional spin behavior over a substantial region of the $k_x$–$k_y$ plane (see Fig. \ref{fig:spintextureGamma}(e,g)). 

\begin{figure}
\includegraphics[width=\linewidth]{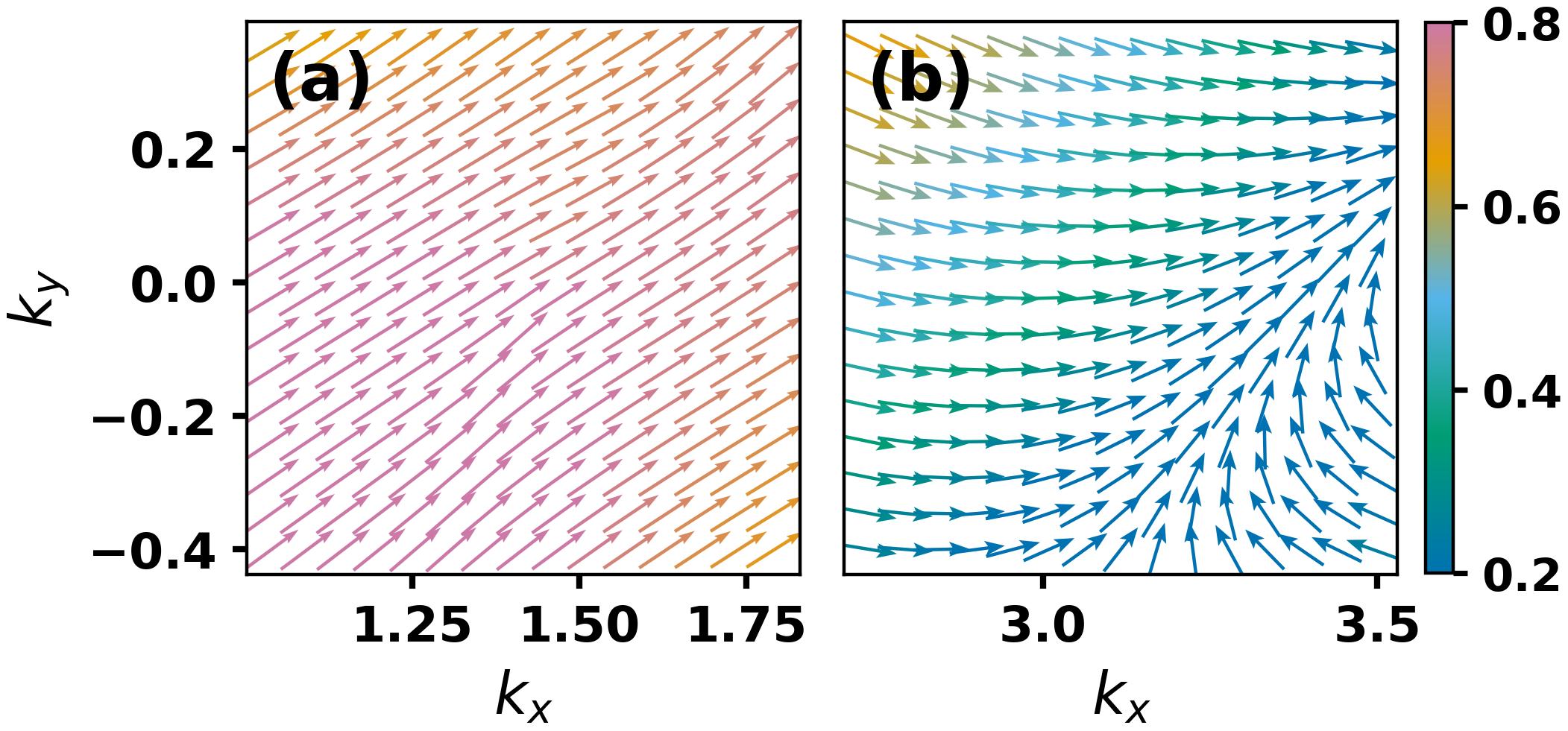}
\caption{Spin textures around $P (1.44, 0.00)$ and $R (3.14, 1.05)$ points for $\eta=0$ and $\eta=1$, respectively. It is clear that although the spins  are partly unidirectional, the magnitudes varies substantially.}
\label{fig:spintexturePQ}
\end{figure}
\subsubsection {Approximate effective Hamiltonian  around $X=(\pi, 0)$}
Around $X=(\pi,0)$, the effective Hamiltonian has the following form  
\begin{align}
H_{\rm low}^{X}(q_x,q_y)
&=
\Big[
r_0(\eta_2-\eta_1)
+s_2 q_x
+s_3 q_y
\Big]\sigma_z
\nonumber\\
&\quad+
\Big[
r_1(\eta_1-\eta_2)
+r_2 q_x
+r_3 q_y
\Big]\sigma_x
\nonumber\\
&\quad+
\Big[
r_1'(\eta_1-\eta_2)
+r_2' q_x
+r_3' q_y
\Big]\sigma_y .
\end{align}
where $r,r_0,r_1,r_2,r_3,r_1',r_2',r_3'$ are functions of the parameters $v,w,\gamma,\delta,\tau,\alpha$, and $\zeta$  (see Appendix \ref{app:X}).

For $\eta_1=\eta_2$, the leading-order contributions to all Pauli matrices are momentum dependent, similar to the case at $\Gamma$. Consequently, a strong momentum-dependent spin texture is expected around the $X$ point, as corroborated by Fig.~\ref{fig:spintextureGamma} (c,d). In contrast, for $\eta_1\neq\eta_2$ with $\eta_1\gg\eta_2$, the leading-order terms associated with all Pauli matrices become momentum independent. This directly results in a persistent spin structure, as clearly evident from Fig.~\ref{fig:spintextureGamma}(f,h). We note that the low-energy effective Hamiltonian around $M$ are found to have similar structure as $X$ and we do not present here for better readability. The effective Hamiltonians around $P$ and $R$ can be obtained using the above expansion method. It turns out that both $H_{\rm low}^{P}(q_x,q_y)$ and $H_{\rm low}^{R}(q_x,q_y)$ contain momentum-dependent leading order corrections to all the Pauli matrices with $\eta_1=\eta_2$, corroborating the results shown in Fig.~(\ref{fig:spintexturePQ}). This analysis elucidates how imaginary hopping can give rise to unconventional spin configurations in a nontrivial topological background, as discussed earlier. Thus, the emergence of persistent spin textures in the present construction differs fundamentally from the conventional mechanisms that rely on a fine-tuned balance between Rashba and Dresselhaus spin–orbit couplings or on nonsymmorphic space-group symmetries\cite{junyi}.   %
	
\section{Summary and Discussion}
	In this work, we have introduced a two-dimensional quasi Su–Schrieffer–Heeger (SSH) model that supports distinct spin-resolved topological phases arising from the interplay of complex hopping and Rashba spin–orbit coupling. These phases are (i) the quantum anomalous spin Hall (QASH) phase, where only a single spin channel carries nontrivial topology with same or opposite chirality and (ii) the quantum anomalous Hall (QAH) phase, in which both spin sectors share the same Chern number. Wherever possible, we have derived approximate low-energy Dirac Hamiltonians to elucidate the microscopic origin of the distinct topological phases. Our analysis reveals how the interplay between complex hopping and spin–orbit coupling gives rise to nontrivial phases. Additionally, we discuss persistent spin textures embedded within a topological background, a feature that is typically absent in conventional topological models due to strong spin mixing induced by spin–orbit interactions. By constructing effective Hamiltonians near high-symmetry points of the Brillouin zone, we discuss the emergence of these nontrivial spin textures. Taken together, these results provide a unified framework that connects topological phases and the evolution of spin textures, thereby clarifying how complex hopping and Rashba-type spin–orbit interactions shape spin-resolved topology in two-dimensional systems.

The proposed lattice model can be realized using ultracold atoms in a two-dimensional optical lattice with Raman-assisted tunneling. A square optical lattice generated by two orthogonal standing-wave laser fields creates periodic potentials along the $x$ and $y$ directions. By introducing an additional superlattice potential along the $x$ direction, each unit cell can be engineered to contain two inequivalent sites $A$ and $B$, forming an SSH-type geometry \cite{Atala2013}. The natural tunneling between these sites gives rise to the intracell hopping amplitude $v$.

Complex tunneling amplitudes can be engineered by applying a linear potential gradient that suppresses the natural hopping between neighboring sites and restoring it via Raman-assisted tunneling. In this scheme, pairs of Raman laser beams induce resonant tunneling between adjacent lattice sites while imprinting a controllable Peierls phase, resulting in an effective hopping amplitude $t_{\mathrm{eff}} = t e^{i\phi}$, where $\phi$ is determined by the momentum transfer of the Raman beams \cite{Aidelsburger2013,Miyake2013}. By tuning the Raman beam phases such that $\phi = \pi/2$, purely imaginary hopping amplitudes can be realized. In particular, Raman-assisted tunneling along the $x$ direction can generate complex intercell hoppings of the form $i\eta_1\, b_{n,m}^\dagger a_{n+1,m}$ and $i\eta_2\, a_{n,m}^\dagger b_{n+1,m}$, while an additional pair of Raman beams with momentum transfer along the $y$ direction can produce the transverse hopping $i\tau\, c_{n,m}^\dagger c_{n,m+1}$. Such control over tunneling phases has been widely demonstrated in experiments realizing synthetic gauge fields and Hofstadter-type Hamiltonians in optical lattices \cite{Aidelsburger2013,Goldman2014}. The model can be further extended to include Rashba spin--orbit interaction (RSOI) using Raman-induced spin--momentum coupling. In this approach, two hyperfine states of the atoms serve as effective spin degrees of freedom, and Raman laser beams couple these states while imparting momentum to the atoms \cite{Lin2011,Galitski2013}. This mechanism generates synthetic spin--orbit coupling that mimics Rashba-type interactions in solid-state systems.

%The proposed model is well suited to realization across a variety of modern quantum platforms. In cold-atom optical lattices, laser-assisted tunneling enables precise control over complex hopping amplitudes, while synthetic spin–orbit coupling can be engineered via Raman coupling schemes. Photonic and phononic lattices provide an alternative route, where complex hopping naturally emerges through engineered phase delays and lattice geometry, and spin–orbit–like effects can be simulated using polarization or mode degrees of freedom. In solid-state settings, one-dimensional chains formed at surfaces or interfaces of materials with strong spin–orbit coupling—such as topological insulators or heavy-element semiconductors—offer a natural platform where both complex hopping and spin–orbit interactions arise intrinsically.

\section{acknowledgment}
KS thanks  Warlley H Camposwa and L. Smejkal for useful discussion. HKS and KS acknowledge financial support from the Department of Atomic Energy (DAE), Govt. of India, through the project Basic Research in Physical and Multidisciplinary Sciences via RIN4001. KS acknowledges funding from the Science and Engineering Research Board (SERB) under SERB-MATRICS Grant No. MTR/2023/000743.

%\onecolumngrid

\appendix
	\section{Effective Spin Hamiltonian}
	\label{app:dirac_derivation}
In this appendix we derive the low-energy effective Hamiltonian around the $\Gamma$ point and establish the origin of the coefficients appearing in the main text. The derivation has two main goals:

(i) to show how the original four-band spin--orbit-coupled Hamiltonian reduces to an effective two-band spin Hamiltonian, and  

(ii) to identify explicitly the momentum-independent and momentum-linear spin terms responsible for the unconventional spin texture discussed in the main text.

Throughout, we retain only the leading contributions relevant to the low-energy physics near $\Gamma$ and $X$.

% ------------------------------------------------
\subsection*{1. Momentum-space Hamiltonian}

The full Hamiltonian in momentum space is written as
\begin{align}
H=\sum_{\mathbf{k}}
\Big[
2\tau \sin k_y\, \tau_z \otimes \mathbb{I}_2
+\tau^+ \otimes M(\mathbf{k})
+\tau^- \otimes M^\dagger(\mathbf{k})
\Big].
\end{align}

Here $\tau_i$ denote Pauli matrices acting in orbital (sublattice) space, while $\sigma_i$ act in spin space. The Hamiltonian therefore lives in the tensor-product space of orbital and spin degrees of freedom.

The inter-orbital coupling matrix is
\begin{align}
M(\mathbf{k})
=
A(\mathbf{k})\mathbb{I}_2
+B(\mathbf{k})\sigma_z
+D_x(\mathbf{k})\sigma_x
+D_y(\mathbf{k})\sigma_y .
\end{align}

The coefficients have the following interpretation:

$A(\mathbf{k})$ spin-independent hopping,
 $B(\mathbf{k})$ spin-dependent term along $\sigma_z$,
 $D_x(\mathbf{k})$ and $D_y(\mathbf{k})$: spin--orbit coupling terms generating in-plane spin mixing.

Their explicit forms are
\begin{align}
A(\mathbf{k}) &= v
+(w-i\eta_1)e^{-ik_x}
+(\gamma+i\eta_2)e^{ik_x}\nonumber\\
&+2\delta\cos k_y(1-e^{-ik_x}), \\
B(\mathbf{k}) &= i\zeta(1-e^{-ik_x}), \\
D_x(\mathbf{k}) &= 2\alpha(\sin k_y+i\cos k_y e^{-ik_x}), \\
D_y(\mathbf{k}) &= -i\alpha(1-e^{-ik_x})
+2\alpha e^{-ik_x}\sin k_y .
\end{align}

The asymmetry between $\eta_1$ and $\eta_2$ will be shown to generate a finite spin term already at zero momentum. In the orbital basis the Hamiltonian takes the block form
\begin{align}
H=
\begin{pmatrix}
H_{\alpha\alpha} & H_{\alpha\beta} \\
H_{\beta\alpha} & H_{\beta\beta}
\end{pmatrix},
\end{align}
with
\begin{align}
H_{\alpha\alpha} &= 2\tau \sin k_y\, \mathbb{I}_2, \\
H_{\beta\beta} &= -2\tau \sin k_y\, \mathbb{I}_2,
\end{align}
and
\begin{align}
H_{\alpha\beta}
=
A\mathbb{I}_2
+D_x\sigma_x
+D_y\sigma_y
+B\sigma_z,
\qquad
H_{\beta\alpha}=H_{\alpha\beta}^\dagger.
\end{align}

Introducing the vector
\begin{align}
\vec D=(D_x,D_y,B),
\end{align}
the off-diagonal block can be written compactly as
\begin{align}
H_{\alpha\beta}=A\mathbb{I}_2+\vec D\cdot\vec\sigma .
\end{align}

% -----------------------------------------------------
\paragraph*{Product of the off-diagonal blocks:}

The key quantity entering the Löwdin effective Hamiltonian is
\begin{align}
H_{\alpha\beta}H_{\beta\alpha}
&=
(A\mathbb{I}_2+\vec D\cdot\vec\sigma)
(A^*\mathbb{I}_2+\vec D^{\,*}\cdot\vec\sigma).
\end{align}
Expanding the product gives
\begin{align}
H_{\alpha\beta}H_{\beta\alpha}
&=
|A|^2\mathbb{I}_2
+A(\vec D^{\,*}\cdot\vec\sigma)
+A^*(\vec D\cdot\vec\sigma)\nonumber\\
&+(\vec D\cdot\vec\sigma)
(\vec D^{\,*}\cdot\vec\sigma).
\end{align}

We now employ the Pauli-matrix identity
\begin{align}
(\vec a\cdot\vec\sigma)(\vec b\cdot\vec\sigma)
=
(\vec a\cdot\vec b)\mathbb{I}_2
+i(\vec a\times\vec b)\cdot\vec\sigma ,
\end{align}
which for $\vec a=\vec D$ and $\vec b=\vec D^{\,*}$ yields
\begin{align}
(\vec D\cdot\vec\sigma)(\vec D^{\,*}\cdot\vec\sigma)
=
(\vec D\cdot\vec D^{\,*})\mathbb{I}_2
+i(\vec D\times\vec D^{\,*})\cdot\vec\sigma .
\end{align}

% -----------------------------------------------------
\paragraph*{Separation into scalar and spin parts :}

Collecting terms proportional to the identity matrix gives
\begin{align}
d_0 = |A|^2+\vec D\cdot\vec D^{\,*}.
\end{align}
The spin-dependent contribution becomes
\begin{align}
\vec d\cdot\vec\sigma
=
A\vec D^{\,*}\cdot\vec\sigma
+A^*\vec D\cdot\vec\sigma
+i(\vec D\times\vec D^{\,*})\cdot\vec\sigma .
\end{align}
Hence,
\begin{align}
\vec d
=
A\vec D^{\,*}+A^*\vec D
+i(\vec D\times\vec D^{\,*}).
\end{align}
Using
\begin{align}
A\vec D^{\,*}+A^*\vec D
=
2\,\mathrm{Re}(A\vec D^{\,*}),
\end{align}
one obtains
\begin{align}
\vec d
=
2\,\mathrm{Re}(A\vec D^{\,*})
+i(\vec D\times\vec D^{\,*}).
\end{align}

% -----------------------------------------------------
\paragraph*{Explicit evaluation of the cross product :}

For $\vec D=(D_x,D_y,B)$,
\begin{align}
\vec D\times\vec D^{\,*}
=
\begin{pmatrix}
D_yB^*-BD_y^* \\
BD_x^*-D_xB^* \\
D_xD_y^*-D_yD_x^*
\end{pmatrix}.
\end{align}
Using $X-X^*=2i\,\mathrm{Im}(X)$,
\begin{align}
i(\vec D\times\vec D^{\,*})
=
\begin{pmatrix}
-2\,\mathrm{Im}(D_yB^*) \\
-2\,\mathrm{Im}(BD_x^*) \\
-2\,\mathrm{Im}(D_xD_y^*)
\end{pmatrix}.
\end{align}

Combining all contributions gives
\begin{align}
d_x &= 2\mathrm{Re}(AD_x^*)
      -2\mathrm{Im}(D_yB^*), \\
d_y &= 2\mathrm{Re}(AD_y^*)
      -2\mathrm{Im}(BD_x^*), \\
d_z &= 2\mathrm{Re}(AB^*)
      -2\mathrm{Im}(D_xD_y^*).
\end{align}

% -----------------------------------------------------
\paragraph*{Effective Hamiltonian :}

Since the Löwdin partitioning yields
\begin{align}
H_{\rm eff}
=
2\tau\sin k_y\,\mathbb{I}_2
+\frac{1}{E+2\tau\sin k_y}
H_{\alpha\beta}H_{\beta\alpha},
\end{align}
the effective spin field components become
\begin{align}
d_x &= \frac{2}{E+2\tau\sin k_y}
\Big[\mathrm{Re}(AD_x^*)-\mathrm{Im}(D_yB^*)\Big], \\
d_y &= \frac{2}{E+2\tau\sin k_y}
\Big[\mathrm{Re}(AD_y^*)-\mathrm{Im}(BD_x^*)\Big], \\
d_z &= \frac{2}{E+2\tau\sin k_y}
\Big[\mathrm{Re}(AB^*)-\mathrm{Im}(D_xD_y^*)\Big].
\end{align}

Therefore the effective Hamiltonian assumes the generic form
\begin{align}
H_{\rm eff}
=
d_0\sigma_0+d_x\sigma_x+d_y\sigma_y+d_z\sigma_z,
\end{align}
where $\mathbf d(\mathbf k)=(d_x,d_y,d_z)$ acts as an
effective momentum-dependent magnetic field governing the
spin texture..

% =====================================================
\section{Expansion around the $\Gamma$ point}
\label{app:Gamma}
We now derive the low-energy form of the effective Hamiltonian
by expanding around the $\Gamma$ point,
\begin{align}
k_x=q_x,\qquad k_y=q_y,
\qquad |q_x|,|q_y|\ll1 .
\end{align}
To linear order in momentum we use
\begin{align}
e^{\pm iq_x} &\approx 1 \pm i q_x, \\
\sin q_y &\approx q_y, \\
\cos q_y &\approx 1 .
\end{align}

% -----------------------------------------------------
\paragraph*{Expansion of the hopping amplitudes:}

Starting from
\begin{align}
A(\mathbf k)
&= v+(w-i\eta_1)e^{-ik_x}
+(\gamma+i\eta_2)e^{ik_x}\nonumber\\
&+2\delta\cos k_y(1-e^{-ik_x}),
\end{align}
we substitute the small-momentum expansions:
\begin{align}
A
&= v+(w-i\eta_1)(1-iq_x)
+(\gamma+i\eta_2)(1+iq_x)\nonumber\\
&+2\delta(1-(1-iq_x)).
\end{align}
Collecting constant and linear terms yields
\begin{align}
A
&= a_0+i(\eta_2-\eta_1)
-(\eta_1+\eta_2)q_x
+i(-w+\gamma+2\delta)q_x ,
\end{align}
where
\begin{align}
a_0=v+w+\gamma .
\end{align}

The remaining coefficients become
\begin{align}
B &= i\zeta(1-e^{-iq_x})
      \approx -\zeta q_x, \\
D_x
&=2\alpha(\sin q_y+i\cos q_y e^{-iq_x})
   \approx 2i\alpha+2\alpha(q_x+q_y), \\
D_y
&=-i\alpha(1-e^{-iq_x})
  +2\alpha e^{-iq_x}\sin q_y
  \approx \alpha q_x+2\alpha q_y .
\end{align}

Thus, to linear order,
\begin{align}
A &= A_0 + A_1 q_x, \\
A_0 &= a_0+i(\eta_2-\eta_1), \\
A_1 &= -(\eta_1+\eta_2)
+i(-w+\gamma+2\delta).
\end{align}

The effective spin field is defined as
\begin{align}
d_x &= \frac{2}{\Lambda}
\Big[\mathrm{Re}(AD_x^*)-\mathrm{Im}(D_yB^*)\Big],\\
d_y &= \frac{2}{\Lambda}
\Big[\mathrm{Re}(AD_y^*)-\mathrm{Im}(BD_x^*)\Big],\\
d_z &= \frac{2}{\Lambda}
\Big[\mathrm{Re}(AB^*)-\mathrm{Im}(D_xD_y^*)\Big],
\end{align}
with
\begin{align}
\Lambda = E+2\tau\sin k_y
\approx E+2\tau q_y .
\end{align}
To linear order we replace $\Lambda\approx E$.

% -----------------------------------------------------
\paragraph*{(i) Expansion of $d_x$ :}

Using
\begin{align}
D_x^*=-2i\alpha+2\alpha(q_x+q_y),
\end{align}
one obtains
\begin{align}
\mathrm{Re}(AD_x^*)
&=
2\alpha(\eta_1-\eta_2)
+2\alpha a_0(q_x+q_y).
\end{align}
The term $\mathrm{Im}(D_yB^*)$ is of order $q^2$ and
is therefore neglected. Hence
\begin{align}
d_x
=
\frac{4\alpha}{E}
\Big[
(\eta_1-\eta_2)
+a_0(q_x+q_y)
\Big].
\end{align}

% -----------------------------------------------------
\paragraph*{(ii) Expansion of $d_y$ :}

Since $D_y$ is already linear in momentum,
\begin{align}
\mathrm{Re}(AD_y^*)
=
a_0(\alpha q_x+2\alpha q_y).
\end{align}
The second term contributes
\begin{align}
-\mathrm{Im}(BD_x^*)
=2\alpha\zeta q_x .
\end{align}
Therefore
\begin{align}
d_y
=
\frac{2}{E}
\Big[
\alpha a_0 q_x
+2\alpha a_0 q_y
+2\alpha\zeta q_x
\Big].
\end{align}

% -----------------------------------------------------
\paragraph*{(iii) Expansion of $d_z$ :}

Since $B\propto q_x$,
\begin{align}
\mathrm{Re}(AB^*)
=-a_0\zeta q_x .
\end{align}
Furthermore,
\begin{align}
-\mathrm{Im}(D_xD_y^*)
=4\alpha^2 q_y .
\end{align}
Hence
\begin{align}
d_z
=
\frac{2}{E}
\Big[
- a_0\zeta q_x
+4\alpha^2 q_y
\Big].
\end{align}

\subsection*{Compact Form of the Low--Energy Hamiltonian at $\Gamma$}

Collecting all linear terms, the effective Hamiltonian near
$\Gamma$ takes the form
\begin{align}
H_{\rm low}^{\Gamma}(q_x,q_y)
&=(f_1' q_x+f_2' q_y)\sigma_y
+(f_0' q_y +f_3 q_x)\sigma_z \nonumber\\
&\quad+
\big[
f_0(\eta_1-\eta_2)
+f_1 q_x
+f_2 q_y
\big]\sigma_x ,
\end{align}
with the coefficients explicitly given by
\[
f_0=\frac{4\alpha}{E},\quad
f_1=\frac{4\alpha a_0}{E},\quad
f_2=\frac{4\alpha a_0}{E},\quad
f_1'=\frac{2\alpha (a_0+2\zeta)}{E},\]\[
f_2'=\frac{4\alpha a_0}{E},\quad
f_0'=\frac{8\alpha^2}{E},\quad
f_3=-\frac{2 a_0 \zeta}{E}.
\]

% -----------------------------------------------------

The term
\begin{align}
f_0(\eta_1-\eta_2)\sigma_x
\end{align}
is momentum independent and survives exactly at the
$\Gamma$ point.
%------------------------------------------------
% 
	%====================================================
\section{Low--Energy Expansion Around the $X=(\pi,0)$ Point}
\label{app:X}
In this section we derive the effective low--energy Hamiltonian
around the high-symmetry point
\(
X=(\pi,0)
\).
We introduce small deviations
\begin{align}
k_x=\pi+q_x,\qquad
k_y=q_y,
\qquad |q_x|,|q_y|\ll1 ,
\end{align}
and retain only terms up to linear order in momenta.

Using standard small-momentum expansions, the relevant Bloch
factors become
\begin{align}
e^{-ik_x} &= -e^{-iq_x}\approx -1+i q_x, \\
e^{ik_x} &= -e^{iq_x}\approx -1-i q_x, \\
\sin k_y &\approx q_y, \\
\cos k_y &\approx 1 .
\end{align}
These relations determine the linearized form of the hopping
amplitudes near the $X$ point.

%---------------------------------------------------
Substituting the above expansions into the microscopic model,
the coefficient $A(\mathbf{k})$ takes the form
\begin{align}
A =
a_X+i\Delta\eta
+(\eta_1+\eta_2)q_x
+i\lambda q_x ,
\end{align}
where
\begin{align}
a_X &= v-w-\gamma+4\delta, \\
\Delta\eta &= \eta_1-\eta_2, \\
\lambda &= w-\gamma-2\delta .
\end{align}
Here $a_X$ represents the effective static mass term at $X$,
while $\Delta\eta$ controls the non-Hermitian imbalance.

The remaining coefficients become
\begin{align}
B &= 2i\zeta+\zeta q_x, \\
D_x &= -2i\alpha+2\alpha(q_y-q_x), \\
D_y &= -2i\alpha-\alpha q_x-2\alpha q_y .
\end{align}
The constant imaginary parts originate from spin--orbit coupling,
whereas the linear terms encode momentum-dependent corrections.

%----------------------------------------------------
\subsection*{Effective Low--Energy Hamiltonian}

Projecting onto the low-energy subspace yields the effective
two-band Hamiltonian
\begin{align}
H_{\mathrm{low}}^{X}
=
d_x\sigma_x+d_y\sigma_y+d_z\sigma_z ,
\end{align}
with coefficients
\begin{align}
d_x &= \frac{2}{E}
\left[\mathrm{Re}(AD_x^*)
-\mathrm{Im}(D_yB^*)\right], \\
d_y &= \frac{2}{E}
\left[\mathrm{Re}(AD_y^*)
-\mathrm{Im}(BD_x^*)\right], \\
d_z &= \frac{2}{E}
\left[\mathrm{Re}(AB^*)
-\mathrm{Im}(D_xD_y^*)\right].
\end{align}

Below we summarize the resulting expressions after keeping only
linear terms in $(q_x,q_y)$.

%----------------------------------------------------
\paragraph*{Coefficient $d_x$:}

Using
\(
D_x^*=2i\alpha+2\alpha(q_y-q_x)
\),
we obtain
\begin{align}
\mathrm{Re}(AD_x^*)
&=
-2\alpha\Delta\eta
+2\alpha a_X(q_y-q_x)
-2\alpha\lambda q_x ,
\\
\mathrm{Im}(D_yB^*)
&=4\alpha\zeta q_y .
\end{align}
Hence
\begin{align}
d_x
=
\frac{4\alpha}{E}
\Big[
-\Delta\eta
+a_X(q_y-q_x)
-\lambda q_x
-2\zeta q_y
\Big].
\end{align}

%----------------------------------------------------
\paragraph*{Coefficient $d_y$:}
Similarly, with
\(
D_y^*=2i\alpha-\alpha q_x-2\alpha q_y
\),
we find
\begin{align}
\mathrm{Re}(AD_y^*)
&=
-2\alpha\Delta\eta
-a_X\alpha q_x
-2a_X\alpha q_y
-2\alpha\lambda q_x ,
\\
\mathrm{Im}(BD_x^*)
&=
4\alpha\zeta q_y
-2\alpha\zeta q_x .
\end{align}
The resulting expression is
\begin{align}
d_y
=
\frac{4\alpha}{E}
\Big[
-\Delta\eta
-\frac{a_X}{2}q_x
-a_X q_y
-\lambda q_x
+\zeta q_x
-2\zeta q_y
\Big].
\end{align}

%----------------------------------------------------
\paragraph*{Coefficient $d_z$:}

For the $\sigma_z$ component we obtain
\begin{align}
\mathrm{Re}(AB^*)
&=
2\zeta\Delta\eta
+\zeta(a_X+2\lambda)q_x ,
\\
\mathrm{Im}(D_xD_y^*)
&=
-2\alpha^2 q_x
+8\alpha^2 q_y .
\end{align}
Thus
\begin{align}
d_z
=
\frac{4}{E}
\Big[
\zeta\Delta\eta
+\frac{\zeta}{2}(a_X+2\lambda)q_x
+\alpha^2 q_x
-4\alpha^2 q_y
\Big].
\end{align}
\vspace{.5cm}
%---------------------------------------------------
% 
%---------------------------------------------------
\subsection*{Compact Form of the Low--Energy Hamiltonian at $X$}

Collecting all constant and linear terms in $(q_x,q_y)$,
the effective Hamiltonian near the $X=(\pi,0)$ point
can be written in compact form.
\begin{align}
H_{\rm low}^{X}(q_x,q_y)
&=
\Big[
r_0(\eta_2-\eta_1)
+s_2 q_x
+s_3 q_y
\Big]\sigma_z
\nonumber\\
&\quad+
\Big[
r_1(\eta_1-\eta_2)
+r_2 q_x
+r_3 q_y
\Big]\sigma_x
\nonumber\\
&\quad+
\Big[
r_1'(\eta_1-\eta_2)
+r_2' q_x
+r_3' q_y
\Big]\sigma_y .
\end{align}

Here the coefficients are given by
\begin{align}
&r_0= -\frac{4\zeta}{E},~~~~
r_1= -\frac{4\alpha}{E},~~~~
r_1' = -\frac{4\alpha}{E},
\nonumber\\[4pt]
&r_2= -\frac{4\alpha}{E}(a_X+\lambda),
r_3= \frac{4\alpha}{E}(a_X-2\zeta),
\nonumber\\[4pt]
&r_2'= -\frac{4\alpha}{E}
\left(\frac{a_X}{2}+\lambda-\zeta\right),
r_3'= -\frac{4\alpha}{E}(a_X+2\zeta),
\nonumber\\[4pt]
&s_2= \frac{4}{E}
\left[
\frac{\zeta}{2}(a_X+2\lambda)
+\alpha^2
\right],
s_3= -\frac{16\alpha^2}{E}.
\end{align}

\bibliographystyle{apsrev4-2}
\bibliography{bib}
	% ==================== Appendix ====================
\end{document}